\documentclass[preprint,12pt]{elsarticle}
\usepackage{amsmath,amsfonts}
\usepackage[caption=false,font=normalsize,labelfont=sf,textfont=sf]{subfig}
\usepackage{textcomp}
\usepackage{stfloats}
\usepackage{url}
\usepackage{verbatim}
\usepackage{graphicx}
\usepackage{booktabs}
\allowdisplaybreaks
\journal{Applied Energy}
\begin{document}

\title{Optimizing Software Defined Battery Systems for Transformer Protection}

\author[1]{Sonia Martin\fnref{fn1} \corref{cor1}}
\author[2]{Obidike Nnorom, Jr. \fnref{fn1}}

\author[3]{Philip Levis}
\author[2,4]{Ram Rajagopal}

\affiliation[1]{organization={Department of Mechanical Engineering, Stanford University},
            addressline={440 Escondido Mall}, 
            city={Stanford},
            postcode={94305}, 
            state={CA},
            country={USA}}

\affiliation[2]{organization={Department of Electrical Engineering, Stanford University},
            addressline={350 Jane Stanford Way}, 
            city={Stanford},
            postcode={94305}, 
            state={CA},
            country={USA}}

\affiliation[3]{organization={Department of Computer Science, Stanford University},
            addressline={353 Jane Stanford Way}, 
            city={Stanford},
            postcode={94305}, 
            state={CA},
            country={USA}}

\affiliation[4]{organization={Department of Civil \& Environmental Engineering, Stanford University},
            addressline={472 Via Ortega}, 
            city={Stanford},
            postcode={94305}, 
            state={CA},
            country={USA}}

\begin{abstract}
  Residential electric vehicle charging causes large spikes in 
electricity demand that risk violating neighborhood transformer power limits. Battery energy storage systems reduce these transformer limit violations, but operating them individually is not cost-optimal. Instead of individual optimization, aggregating, or sharing, these batteries leads to cost-optimal performance, but homeowners must relinquish battery control. This paper leverages virtualization to propose battery sharing optimization schemes to reduce electricity costs, extend the lifetime of a residential transformer, and maintain homeowner control over the battery. 
A case study with simulated home loads, solar generation, and electric vehicle charging profiles 
demonstrates that joint, or shared, optimization reduces consumer bills by 56\% and transformer aging by 48\% compared to individual optimization. Hybrid and dynamic optimization schemes that provide owners with autonomy have similar transformer aging reduction but are slightly less cost-effective. These results suggest 
that controlling shared batteries with virtualization is an effective way to delay transformer upgrades in the face of growing residential electric vehicle charging penetration.


  \end{abstract}


\cortext[cor1]{Corresponding Author: soniamartin@stanford.edu}
\fntext[fn1]{Equal contribution}


\maketitle

  \section{Introduction}
Increasing residential electric vehicle (EV) penetration can harm distribution transformers~\cite{hilshey2013estimating}. Since utility companies chose many transformer sizes before widespread installation of residential EV chargers, high peak power from charging can lead to capacity violations and consequently, transformer aging~\cite{muratori2018impact, gong2012study, soleimani2020economic}. This rapid aging forces expensive and time-consuming transformer upgrades~\cite{sarker2017cooptimization}. As of 2023, distribution transformer prices in the U.S. have more than doubled and the installation wait time exceeds one year~\cite{kaufmanRunningLowMachines2023}. As more EV chargers are installed in neighborhoods with limited transformer capacities, it is in the utilities' best interest to implement techniques to prolong transformer lifetime instead of upgrading these assets~\cite{li2024impact}.


While controlling EV charging can help reduce peak loads and protect transformers~\cite{hilshey2013estimating, sarker2017cooptimization, powell2020controlled, li0222ev, affonso2019technical, visakh2022controlled, rossi2025smart, sarmokadam2025power}, drivers may not respond perfectly to charging control signals, which risks unplanned transformer limit violations~\cite{latinopoulos2017response}. Instead, controlling stationary distributed energy resources (DERs), such as stationary battery storage systems (BESSs), offers a solution to protect transformers that is decoupled from drivers' charging behavior~\cite{navidi2023coordinating, endara2025design}. For example, a homeowner with a BESS can take advantage of excess solar photovoltaic energy generated to charge the BESS and then discharge it during an EV charging session to offset peak demand. With this type of peak shaving, BESSs not only reduce peak power costs but also mitigate the large power spikes from EV charging that can harm transformers.

Many studies on BESSs co-located with EVs focus only on controlling a single BESS, with control objectives ranging from transformer protection~\cite{datta2020smart, hong2020assessment} and cost minimization~\cite{li2015aggregator} to voltage control~\cite{moradiamani2023data}. However, a single centrally controlled BESS does not account for individual homeowner needs if implemented in a residential neighborhood or distribution network~\cite{raghuveer2025real}. In regions with frequent power outages or high peak prices, a homeowner buying into a BESS system may want to retain autonomy over their BESS to maintain reserve power.

 Especially in communities, combining, or aggregating, multiple BESS units creates shared infrastructure from which entire neighborhoods can benefit~\cite{ntube2023stochastic, hafiz2019energy, henni2021sharing, lee2021autoshare}.
Aggregation can increase both BESS utilization and decrease consumer costs: correctly sized shared systems that take into account solar and EV system sizes~\cite{khanal2023optimal, merrington2023optimal} have been shown to increase utilization by over 35\% and decrease costs by over 10\%~\cite{walker2021analysis, kang2023multi, barbour2018community, roberts2019impact}. 

However, similar to single BESS control, these aggregated systems are often static. Once a set of batteries is shared, it cannot easily be reverted back to individual control, and vice versa. With unpredictable outages, loads that change seasonally, and variable EV charging patterns, a static aggregation scheme may be suboptimal. In such a system where adaptability is crucial, there is not a consensus on how exactly BESSs should be split among homeowners and how this split should change over time. 

Battery virtualization is a new approach that simplifies operation and allows for more flexibility in the battery setup in the face of variable grid conditions. Virtualization provides an abstraction of a virtual resource that is equivalent to an underlying
physical resource~\cite{lee2021autoshare, martin2022software, bashir2021enabling, souza2023ecovisor}, hiding complexities of the physical resource and making it easier to interact with it. Battery virtualization involves treating each physical battery as an abstract entity that can be partitioned into separate parts or aggregated with other abstract, or software defined, batteries. For example, a group of homes in a neighborhood can use virtualization to combine all of their physical batteries and treat them as a single aggregate battery. Virtualization also enables a hybrid aggregation approach, in which an owner can maintain control of one battery partition while contributing another partition to a shared aggregate. If constituent batteries are added or drop
offline~\cite{lee2021autoshare, martin2022software}, virtualization recalibrates without interruption. Lastly, virtualization helps the system adapt to seasonality changes and uncertainty in forecasts when running a model predictive control (MPC) scheme for BESS control~\cite{moradiamani2023data, kumar2018stochastic, clarke2016economic, zamani2015integration, nair2021analysis}.


Prior work has explored BESS aggregation~\cite{ntube2023stochastic, hafiz2019energy, henni2021sharing, lee2021autoshare} and cost minimization techniques~\cite{li2015aggregator} separately. Combining these two approaches can help protect transformers in the face of increased EV adoption. Flexible and dynamic BESS sharing as well as 
aggregation schemes allowing homeowners to retain autonomy of part of their BESS 
have not yet been investigated for this purpose. 

This paper explores how BESS operators can leverage virtualization software~\cite{martin2022software} to seamlessly control a set of aggregated residential BESSs to both minimize costs and mitigate transformer loss of life while maintaining homeowner autonomy. In contrast with existing work that utilizes static, aggregated BESSs for transformer protection, we compare the performance of multiple static and dynamic aggregation schemes in a case study with simulated load and EV data. Lastly, we test the sensitivity of the algorithm to imperfect forecasting by implementing an MPC scheme and analyzing results across seasons.


We find that introducing flexibly aggregated BESSs reduces expensive transformer upgrades and decreases operational costs. Our proposed algorithms are useful for homeowners, energy aggregators, and utilities aiming to avoid these transformer upgrades, decrease electricity costs, and increase BESS utilization within a residential neighborhood. These cost reductions enable more widespread BESS and EV charger installations, paving the way for cleaner light-duty electric transportation. 

This work makes the following contributions:

\begin{itemize}

       \item Optimization schemes enabled by virtualization for a group of BESSs to reduce costs and transformer aging by sharing the batteries, 
       \item A ``hybrid'' optimization scheme that flexibly combines individual and joint (shared) control using virtualization,
       \item A ``dynamic'' optimization scheme that leverages virtualization to dynamically balance individual and joint control, 
       \item Identification of the best battery sharing fraction to simultaneously reduce costs and maintain homeowner autonomy, and
       \item Analysis of each scheme under imperfect forecasts, with variable pricing schemes, across seasons, with varying EV penetration, and considering battery aging.

\end{itemize}

The paper is organized as follows: Section~\ref{algorithms} explains the optimization schemes, Section~\ref{transformer} describes the transformer model, and Section~\ref{results} presents simulation results. Section~\ref{discussion} discusses the implications of these results while Section~\ref{conclusion} outlines future work.

  \section{Control Algorithms}
  \label{algorithms}
\begin{figure}[ht]
\centering
        \includegraphics[scale=.45]{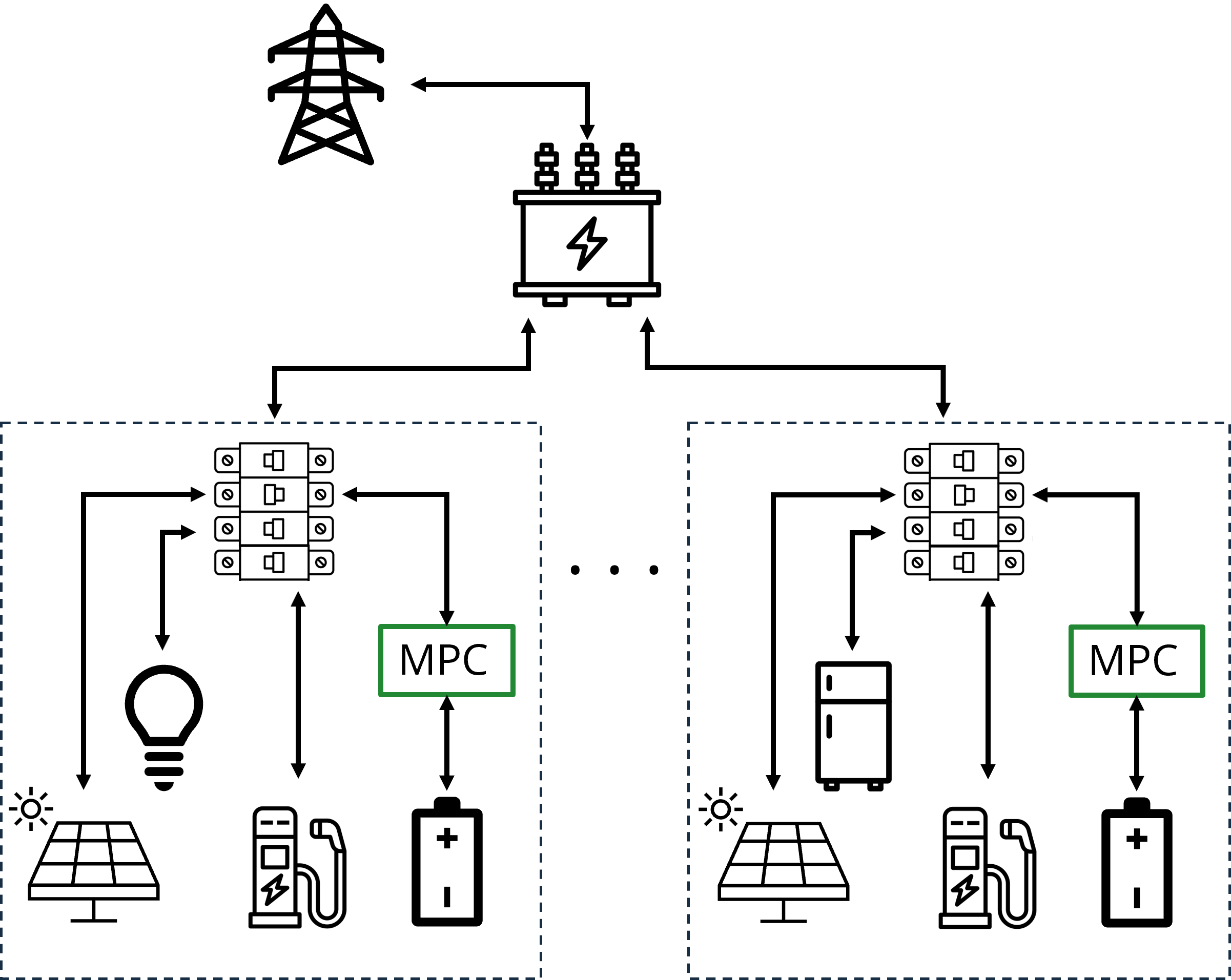}
    \caption{Diagram of a residential power system. A set of homes, connected to the grid under a single transformer, contains home loads, solar, a battery, and an EV charger. 
    Each of these loads is connected to the transformer via a circuit breaker. The model predictive controller determines the battery output.}
    \label{fig:mpc_diagram}
\end{figure}

  This section discusses the optimization strategies, controller, system sizing, forecasting, and electricity cost structure. We consider a setting where a collection of homes in a neighborhood has solar systems, EV chargers, residential BESSs, and various 
home loads. Figure~\ref{fig:mpc_diagram} displays these components for two homes, connected to the distribution grid via a transformer.
This work analyzes the homes downstream of one residential transformer. We assume that solar generation and all 
loads are fixed (i.e. no flexible demand) and that each home's battery is controllable.

We compare five different battery optimization schemes to demonstrate 
the effect of aggregation and virtualization on cost and transformer protection:
\begin{itemize}
    \item \textbf{Individual optimization} assumes that 
each home in a neighborhood has their own battery that they control to minimize their individual cost, 
 \item  \textbf{Individual optimization with uneven transformer allocation} is similar to the first scheme but allows homes to have an uneven limit allocation instead of distributing the limit evenly,
 \item   \textbf{Joint optimization} controls all of the batteries in a neighborhood as if they 
were a single aggregate battery, 
\item \textbf{Hybrid optimization} combines the individual and joint strategies, and 
\item \textbf{Dynamic partitioning optimization} builds off the hybrid scheme by adding dynamic partition values between retained and shared portions of the battery. 
\end{itemize}
We describe the schemes below, with 
variables listed in Table~\ref{tab:variables}.

\begin{table}[ht!]
    \centering
    \begin{tabular}{ll}
     \toprule
     Symbol & Variable Name  \\ 
      \midrule
      $n$                                       & Number of Homes \\
      $t$                                       & Timestep \\
      $T$                                       & Total Number of Timesteps \\
      $\lambda$                                 & Transformer Violation Penalty Weight \\
      $K $                                      & Nominal Transformer Limit [kW]\\
      $\alpha$                                 & (Dis)charging Penalty Weight \\
      $\eta$                                    & One-way Battery Efficiency \\
      $\Delta t$                                & Timestep Duration [h] \\
      $E_{\textrm{max}} $                       & Maximum Battery Capacity [kWh]\\
      $h$                                       & Hours Between Change in Partition Weights \\
      $T^\textrm{MPC}$                          & Number of Timesteps per MPC Iteration \\
      $B_\textrm{max}^\textrm{dischg}, B_\textrm{max}^\textrm{chg} \in \mathbb{R}^+$ & Maximum Battery (Dis)charge Power [kW]\\
      $\mathbf{B}^\textrm{dischg}, \mathbf{B}^\textrm{chg} \in \mathbb{R}^{T \times n}$ & Battery (Dis)charge Power [kW]\\
      $ \mathbf{M} \in \mathbb{R}^{T \times n}$ & Home Meter Reading [kW] \\
      $ \mathbf{C} \in \mathbb{R}^T$            & Electricity Cost [$\frac{\$}{kWh}$]\\
      $ \mathbf{S} \in \mathbb{R}^{T \times n}$ & Home Solar Generation [kW] \\
      $ \mathbf{L} \in \mathbb{R}^{T \times n}$ & Home Load Demand [kW] \\
      $\mathbf{E} \in \mathbb{R}^{(T+1) \times n}$ & Available Battery Capacity [kWh]\\
      $\mathbf{K}^U \in \mathbb{R}^{n}$         & Uneven Transformer Allocation [kW] \\
      $\mathbf{B}^\textrm{R,dischg} \in \mathbb{R}^{T \times n}$ & Retained Discharge Power [kW]\\
      $\mathbf{B}^\textrm{S,dischg} \in \mathbb{R}^{T \times n}$ & Shared Discharge Power [kW]\\
      $\mathbf{B}^\textrm{R,chg} \in \mathbb{R}^{T \times n}$ & Retained Charge Power [kW]\\
      $\mathbf{B}^\textrm{S,chg} \in \mathbb{R}^{T \times n}$ & Shared Charge Power [kW]\\
      $\mathbf{B}^\textrm{R} \in \mathbb{R}^{T \times n}$ & Retained Net Power [kW]\\
      $\mathbf{B}^\textrm{S} \in \mathbb{R}^{T \times n}$ & Shared Net Power [kW]\\
      $\mathbf{W} \in \mathbb{R}^{T \times n}$     & Retained Battery Fraction \\
      $\mathbf{E}^R \in \mathbb{R}^{(T+1) \times n}$ & Retained Battery Capacity [kWh]\\
      $\mathbf{E}^S \in \mathbb{R}^{(T+1) \times n}$ & Shared Battery Capacity [kWh]\\
      $\mathbf{Z} \in \mathbb{R}^{T \times \frac{T \cdot \Delta t}{h}}$ & Dynamic Partitioning Transformation Matrix \\
      $\mathbf{W}^C \in \mathbb{R}^{\frac{T \cdot \Delta t}{h} \times n}$ & Dynamic Partitioning Constant Weights \\
      $C^\textrm{batt}$ & Battery Capital Cost\\ 
      $N^\textrm{cyc}$ & Maximum Cycles under Warranty\\
      \bottomrule
\end{tabular} 
\vspace{.5em}
\caption{Algorithm Variables}
\label{tab:variables}
\end{table}
\clearpage

\subsection{Individual Optimization}

With individual optimization, each homeowner operates their own battery. The neighborhood transformer limit $K$ is split evenly amongst
each of the $n$ homeowners, incentivizing each home's power consumption to remain below $\frac{K}{n}$. Each homeowner 
controls their battery to minimize costs (see tariff information in Section~\ref{subsec:costs}) while ensuring their power consumption does not
exceed the individual transformer limit. The following convex optimization problem solves for the optimal charge and discharge battery output for each home independently:
\begin{align}
    & \underset{\mathbf{B}^\textrm{dischg},~\mathbf{B}^\textrm{chg}}{\min}   & & \sum_{i=0}^n \biggl ( \mathbf{C}^T  \max(\mathbf{M}_i,0)\cdot \Delta t \\
   & & &  + \sum_{t=0}^T \left( \lambda  \max ( \mathbf{M}_{t,i} - \frac{K}{n} , 0)^2 +  \alpha(\mathbf{B}^\textrm{chg}_{t,i} + \mathbf{B}^\textrm{dischg}_{t,i}) \right) \biggr )   \label{eqn:indiv_obj}
\end{align}
\vspace{-1em}
\begin{align}
    & \mathrm{s.t.}  & & \mathbf{M}_{t} =  \mathbf{L}_{t} - \mathbf{S}_{t} +  \mathbf{B}^\textrm{chg}_{t} - \mathbf{B}^\textrm{dischg}_{t}  & \forall t \in [0,T]  \label{eqn:indiv_meter} \\
    &   & & \mathbf{E}_{t+1} = \textbf{E}_{t} - \Delta t  \left(\frac{\mathbf{B}^\textrm{dischg}_{t}}{\eta} - \eta \mathbf{B}^\textrm{chg}_{t} \right) &\forall  t \in [0,T]  \label{eqn:indiv_soc}\\ 
    &   & & 0 \leq \mathbf{E}_{t} \leq E_\textrm{max}  &\forall  t \in [0,T+1] \label{eqn:indiv_soc_limit} \\
    &   & & 0 \leq \mathbf{B}^\textrm{chg}_{t} \leq  B^\textrm{chg}_\textrm{max} & \forall   t \in [0,T]  \label{eqn:indiv_max_dischg_power}  \\
    &   & & 0 \leq \mathbf{B}^\textrm{dischg}_{t} \leq  B^\textrm{dischg}_\textrm{max} & \forall  t \in [0,T]  \label{eqn:indiv_max_chg_power} 
\end{align}

The objective, Equation~\eqref{eqn:indiv_obj}, has three terms: the total electricity cost for the homeowners in the community, 
the transformer violation penalty, and a penalty to discourage the solver from finding solutions in which the battery charges and discharges simultaneously
 (since this is impossible in practice). There is no net metering, so the positive values of the meter are multiplied 
by the timestep length and the cost at each timestep in units of $\frac{\$}{kWh}$. Equation~\eqref{eqn:indiv_meter} 
constrains the utility meter reading to equal the sum of the homeowner's loads and battery output minus their 
solar generation. Equations~\eqref{eqn:indiv_soc},~\eqref{eqn:indiv_soc_limit}, ~\eqref{eqn:indiv_max_dischg_power}, and~\eqref{eqn:indiv_max_chg_power} govern the battery state of charge 
evolution, capacity minimum and maximum, and charge and discharge limits, respectively.

\subsection{Individual Optimization with Uneven Transformer Allocation}

The individual optimization scheme penalizes each homeowner for exceeding their equal allocation of the total transformer limit. In practice, not all homes have the same maximum power draw~\cite{pezeshki2014impact}. This uneven allocation scheme optimizes the fraction of the transformer limit allocated to each home to account for this variability in maximum home power draw.

The scheme has a different objective function and two additional constraints not found in individual optimization, listed in the equations below:
\begin{align}
    & \underset{\mathbf{B}^\textrm{dischg},~\mathbf{B}^\textrm{chg}}{\min}   & &  \sum_{i=0}^n \bigg(\mathbf{C}^T \max(\mathbf{M}_i,0)  \Delta t \nonumber \\
    & & & +\sum_{t=0}^T \left( \lambda  \max(\mathbf{M}_{t,i} - \mathbf{K} ^\mathbf{u}_i , 0)^2 + \alpha(\mathbf{B}^\textrm{chg}_{t,i} + \mathbf{B}^\textrm{dischg}_{t,i}) \right)\bigg) \label{eqn:case_2_obj} \\
    & \mathrm{s.t.}  & & \mathbf{K} ^\mathbf{u}\geq 0 \label{eqn:unevenK_1} \\
    & & &  \sum_{i=0}^n \mathbf{K} ^\mathbf{u}_i = K \label{eqn:unevenK_2}\\
   & & & +\textrm{remaining constraints from individual optimization} \nonumber
\end{align}

Equation~\eqref{eqn:case_2_obj} shows the updated objective function which replaces $K$ in the objective with $\mathbf{K^u}$, a vector 
that represents the transformer limit for each home in kW. Next, Equations~\eqref{eqn:unevenK_1} and~\eqref{eqn:unevenK_2} are added constraints 
to ensure each home has a nonnegative limit and that the sum of these limits over all the homes is equal to the nameplate limit $K$. 
These changes allow the solver to simultaneously find the minimum cost solution and optimal transformer limit allocation.

\subsection{Joint Optimization}

With joint optimization, the homeowners aggregate their batteries together using virtualization~\cite{martin2022software}.
The controller manages the aggregate battery to prevent the homeowners' total power consumption from exceeding the transformer
limit whenever possible. 

The joint scheme allows homeowners to collaborate to protect the transformer. Depending on the exact home energy consumption profiles, the joint scheme can be advantageous compared to the individual scheme: if any given home's battery is underutilized in the individual scheme, another home can use it. The problem is as follows:

\begin{align}
 & \underset{\mathbf{B}^\textrm{dischg},~\mathbf{B}^\textrm{chg}} {\min} & & \mathbf{C}^T\max(\textbf{M}  \mathbf{1} ,0) \cdot \Delta t   +  \sum_{t=0}^T \bigg(  \lambda \max((\textbf{M}  \mathbf{1}) _t - K,0)^2 \label{eqn:obj_joint} \nonumber \\
  & & &  +  \sum_{i=0}^n \alpha (\mathbf{B}_{t,i}^\textrm{chg} + \mathbf{B}_{t,i}^\textrm{dischg}) \bigg)
  \end{align}
\vspace{-1.5em}
\begin{align}
 &\mathrm{s.t.} & &  \mathbf{M}_{t} =  \mathbf{L}_{t} - \mathbf{S}_{t} + \mathbf{B}^\textrm{chg}- \mathbf{B}^\textrm{dischg} & \forall t \in [0,T]   \label{eqn:joint_meter}\\
 & & & \mathbf{E}_{t+1} = \mathbf{E}_t - \Delta t  \left(\frac{\mathbf{B}^\textrm{dischg}_{t}}{\eta} - \eta \mathbf{B}^\textrm{chg}_{t} \right) &  \forall t \in [0,T] \label{eqn:joint_soc}\\ 
 & & & 0 \leq \mathbf{E}_{t} \leq E_{\textrm{max}} & \forall t \in [0,T+1] \label{eqn:joint_soc_limit}\\
 & & & 0 \leq \mathbf{B}^\textrm{chg}_t \leq B^\textrm{chg}_\textrm{max} & \forall  t \in [0,T]  \label{eqn:joint_max_chg_power} \\
 & & & 0 \leq \mathbf{B}^\textrm{dischg}_t \leq  B^\textrm{dischg}_\textrm{max} & \forall  t \in [0,T]  \label{eqn:joint_max_dischg_power}  
\end{align}

The aggregate battery aims to minimize the total homeowners'
cost. A given customer might be restricted below their fair share of power at certain times but can save money by participating in the aggregate.  
The utility bills the aggregate system as a whole: the first term in the objective, 
Equation~\eqref{eqn:obj_joint}, multiplies the sum of all the home meters and the aggregate battery output 
by the TOU cost vector. Similar
to the individual optimization scheme, the second term in Equation~\eqref{eqn:obj_joint} penalizes transformer
limit violations, but here the system is penalized when the sum of the homes' net power and aggregate battery output exceeds the limit, instead of when a an individual home's power peak exceeds its allotted limit. A penalty is also included in the objective to discourage the solver
from choosing to charge and discharge the battery variables simultaneously.  
Equation~\eqref{eqn:joint_meter} constrains each home meter to equal load demand
minus solar generation. Equations~\eqref{eqn:joint_soc},~\eqref{eqn:joint_soc_limit},~\eqref{eqn:joint_max_chg_power} and~\eqref{eqn:joint_max_dischg_power} govern the battery SOC 
evolution, capacity minimum and maximum, and charge and discharge limits, respectively.
 
\subsection{Hybrid Optimization}
In the hybrid scheme, each homeowner keeps a certain fraction of their battery and gives up the remaining fraction to be aggregated into a larger battery. This allows the homeowner to participate in joint optimization while maintaining control of some of their battery, e.g. for blackouts. These fractions, or partitions, are denoted by $\mathbf{E^R}$ and $\mathbf{E^S}$ for the retained and shared partitions, respectively, and weighted by an input $\mathbf{W}$ between 0 and 1. In this scheme, these partitions are fixed for the whole time horizon. Virtualization enables this combination of individual optimization and joint optimization by treating the shared partitions as one virtual battery and each retained partition as additional virtual batteries.

The objective of the hybrid scheme includes terms for the individual, joint, and transformer violation objectives. 
This balances the homeowner's goal to act within their financial interest while also sharing 
part of their battery in the interest of the entire community. The problem is as follows:

\begin{align}
    & \underset{\mathbf{B}^\textrm{R}, \mathbf{B}^\textrm{S}}{\min}   & &  \sum_{i=0}^n \mathbf{C}^T \max\left(\mathbf{M}_i +  \mathbf{B}^\textrm{R}_i ,0\right) \cdot \Delta t  \nonumber \\
     & & &   + \sum_{t=0}^T \bigg(  \mathbf{C}^T \max\left( \left(\textbf{M}\mathbf{1} + \mathbf{B}^\textrm{S}\mathbf{1}\right)_{t}  ,0\right)^2 \cdot \Delta t \nonumber \\
    & & &  +  \lambda  \max\left( \left(\textbf{M}\mathbf{1} + \mathbf{B}^\textrm{S}\mathbf{1}\right)_t  - K,0\right)^2  \nonumber \\
    & & &  +  \sum_{i=0}^n\alpha \left(\mathbf{B}_{t,i}^\textrm{R} + \mathbf{B}_{t,i}^\textrm{S}\right) \bigg) \label{eqn:hybrid_obj} 
      \end{align}
\vspace{-1.5em}
\begin{align}
    & \mathrm{s.t.}  & & \mathbf{M}_{t} =  \mathbf{L}_{t} - \mathbf{S}_{t} & \forall t \in [0,T]  \label{eqn:hybrid_meter} \\
    &   & & \mathbf{B}^\textrm{R} = \mathbf{B}^\textrm{R,chg} - \mathbf{B}^\textrm{R,dischg} \label{eqn:retain} \\
    &   & & \mathbf{B}^\textrm{S} = \mathbf{B}^\textrm{S,chg} - \mathbf{B}^\textrm{S,dischg} \label{eqn:shared} \\
    &   & & \mathbf{E}_{t+1}^R = \textbf{E}_{t}^R - \Delta t \left(\eta \mathbf{B}_{t}^\textrm{R,chg} - \frac{\mathbf{B}_{t}^\textrm{R,dischg}}{\eta} \right) &   \forall  t \in [0,T]  \label{eqn:hybrid_soc_R}\\ 
    &   & & \mathbf{E}_{t+1}^S = \textbf{E}_{t}^S - \Delta t \left(\eta \mathbf{B}_{t}^\textrm{S,chg} - \frac{\mathbf{B}_{t}^\textrm{S,dischg}}{\eta} \right)   & \forall  t \in [0,T]  \label{eqn:hybrid_soc_S}\\ 
    &   & & 0 \leq \mathbf{E}_{t}^R \leq \mathbf{W}_t E_\textrm{max}   & \forall t \in [0,T+1] \label{eqn:hybrid_soc_limit_R} \\
    &   & & 0 \leq \mathbf{E}_{t}^S \leq (1-\mathbf{W}_t) E_\textrm{max} & \forall t \in [0,T+1] \label{eqn:hybrid_soc_limit_S} \\
    &   & & 0 \leq \mathbf{B}^\textrm{R,chg}_{t} \leq \mathbf{W}_t B^\textrm{chg}_\textrm{max}\quad  & \forall  t \in [0,T]  \label{eqn:hybrid_chg_limit_R} \\
    &   & & 0 \leq \mathbf{B}^\textrm{S,chg}_{t} \leq (1-\mathbf{W}_t)  B^\textrm{chg}_\textrm{max} &  \forall  t \in [0,T]  \label{eqn:hybrid_chg_limit_S} \\
    &   & & 0 \leq \mathbf{B}^\textrm{R,dischg}_{t} \leq \mathbf{W}_t B^\textrm{dischg}_\textrm{max} & \forall  t \in [0,T]  \label{eqn:hybrid_dischg_limit_R} \\
    &   & & 0 \leq \mathbf{B}^\textrm{S,dischg}_{t} \leq (1-\mathbf{W}_t) B^\textrm{dischg}_\textrm{max} &   \forall  t \in [0,T]  \label{eqn:hybrid_dischg_limit_S} 
\end{align}

The first term in Equation~\eqref{eqn:hybrid_obj} mirrors the objective of the individual optimization scheme, while the second term is similar to the objective for joint optimization. The transformer penalty only occurs if the home loads, solar, and shared battery exceed the limit, i.e. the retained portion does not factor into the penalty, giving homeowners complete autonomy over it. The constraints here, Equations~\eqref{eqn:hybrid_meter}, \eqref{eqn:hybrid_soc_R}, \eqref{eqn:hybrid_soc_S}, \eqref{eqn:hybrid_soc_limit_R}, \eqref{eqn:hybrid_soc_limit_S}, \eqref{eqn:hybrid_chg_limit_R}, \eqref{eqn:hybrid_chg_limit_S}, \eqref{eqn:hybrid_dischg_limit_R}, and \eqref{eqn:hybrid_dischg_limit_S}, are also similar to both prior schemes, but they are duplicated for each of the retained and shared partitions. The charge/discharge and energy constraints ensure each partition is only allotted $\mathbf{W}$ or $1-\mathbf{W}$ of the power and energy limits. Equations~\eqref{eqn:retain} and \eqref{eqn:shared} set the net battery power equal to the charge minus the discharge power for each partition.

\subsection{Dynamic Partitioning}
While the hybrid scheme takes a static value $\mathbf{W}$, virtualization allows for this weight to change over time. For example, if the occupants of one home go on vacation for a week then return, it is suboptimal if their retained battery fraction remains constant over a whole month. In traditional, static optimization, this fraction is fixed over time. However, with virtualization, this fraction can be dynamic as the virtualization software handles the re-partitioning of the batteries while respecting battery power constraints. We test a case in which $\mathbf{W}$ is allowed to vary every two hours. The optimization problem is the same as in the hybrid case, with the following additional constraints:
\begin{align}
    0 \leq \textbf{W} \leq 1  \label{eqn:alpha_restrict}\\
    \textbf{W} = \mathbf{Z} \textbf{W}^C \label{eqn:alpha_fixed}
\end{align}

Equation~\eqref{eqn:alpha_restrict} restricts $\mathbf{W}$ to between 0 and 1, while Equation~\eqref{eqn:alpha_fixed} (using h = 2) ensures that $\mathbf{W}$ can only change once every two hours.

\subsection{Model Predictive Control}
While a convex optimization solver can easily solve the above problems, the battery output is only optimal
given perfect foresight of the solar, load, and EV charging profiles. In real time operation, these profiles are not known. We use moving horizon 
MPC that updates initial values if the system deviates from its forecast. 
At each MPC iteration, the controller solves for the next day given solar and load forecasts and 
outputs the desired battery trajectory.
Then, the controller selects the initial battery power command and sends this value to the battery. After one timestep, 
the next iteration begins with a new initial battery SOC that is read from the battery. This process repeats until the 
end of the overall optimization horizon. With this moving window, the controller corrects for changes in the system, either due to 
deviations between forecasts and actual values or experimental factors such as sensor bias or error. In the results section, we compare the MPC
with a perfect foresight controller, which is a one-shot optimization over the entire time horizon.

\subsection{System Sizing} \label{violations}
In simulation, we model the batteries as the Tesla Powerwall 2. 
We assume each homeowner owns a single Tesla Powerwall 2 with a maximum capacity of 13.5 kWh and a maximum 
(dis)charge output of 5 kW~\cite{teslapowerwall}. We use a fixed transformer size of 25 kVA as it represents a 
transformer size commonly found in residential communities~\cite{muratori2018impact}.

We select the number of homes located downstream of the transformer based on the home loads. Since older transformers were typically sized before high prevalence of EV chargers, we adopt a heuristic to select a realistic number of homes under each transformer. The heuristic proceeds as follows. First, we randomly sample homes from the dataset. Next, we add these sampled homes, one at a time, until the maximum sum of their loads (excluding EV charging) over a month exceeds the nominal transformer limit; this is called a ``violation''. This set of homes forms the neighborhood behind the transformer for a particular trial.


\subsection{Battery Aging Considerations}
Although a complex battery aging model is out of the scope for this work, we examine the sensitivity of the total system costs to battery aging. Specifically, we add a term to the objective function that represents a cost in \$ per roundtrip charge and discharge cycle. This cost per cycle is derived from the cost of purchasing a new Powerwall, $C^\textrm{batt}$~\cite{TeslasPowerwall2} divided by the number of total cycles allowed under warranty~\cite{teslaTESLAPOWERWALLLIMITED2025}. This represents the effective cost per cycle for the Powerwall. The objective term is as follows in Equation~\eqref{eq:aging}:

\begin{align}
    \frac{C^\textrm{batt}}{N^\textrm{cyc}} \frac{\left( \sum_{t=0}^T \mathbf{B}^\textrm{chg}_t + \sum_{t=0}^T \mathbf{B}^\textrm{dischg}_t \right)\Delta t}{2E^\textrm{max} } \label{eq:aging}
\end{align}

\subsection{Forecasting}
We use a naive forecaster that provides the solar generation and home load demand (excluding EVs) at each home for the 
upcoming day by averaging values from the previous four days. The naive forecaster is inaccurate for EV loads as charging 
times for EVs can be short and somewhat sporadic. As a result, we assume that the EV charging profiles are known in advance. This assumes that homeowners will know one day in advance (the MPC time horizon) when they plan to charge their vehicle.

\subsection{Costs} \label{subsec:costs}
For the main simulations, we adopt the Pacific Gas \& Electric's (PG\&E) Home Charging EV2-A 
time-of-use (TOU) 
rate plan as a tariff structure for each home~\cite{pgeResidentialRatePlan2024}. The pricing scheme, 
effective as of April 2024, is representative of the costs for homeowners located in San Jose with Level 2 EV charging. We also run a simulation that uses the EV-B and TOU-D PG\&E rate structures~\cite{pgeResidentialRatePlan2024}.
For all schemes, we analyze the total costs of a community of homes using the same tariff structure. We do not determine the individual cost distribution for each home as the exact methods used for distributing the costs are outside the scope of this paper. This cost distribution would factor in homes' baseline loads, EV charger utilization, time of peak loads, to name a few inputs. Lastly, this work only considers operational costs of the system and does not factor in capital costs of purchasing a BESS, a solar array, or an EV charger.

  \section{Transformer Model}
  \label{transformer}
  
We model the lifetime and aging of an oil-type, ONAN (natural cooling) residential distribution transformer using equations from the IEEE C57.96 
standard~\cite{IEEEGuideLoading2012}. We focus on oil-type transformers rather than dry-type
transformers as they are common in residential settings~\cite{hilshey2013estimating}. We obtain the model parameters for the 25 kVA low voltage transformer, such as 
power loss at 100\% load and weight of the coils, from a datasheet~\cite{LarsonElectronics25}. As power flow analysis is outside the scope of this work, we assume a power factor of 1 
and thus treat the nominal 25 kVA limit as 25 kW. Table~\ref{tab:transformer_aging_variables} lists the variables in the transformer aging equations.
\begin{table}[ht]
    \centering
    \begin{tabular}{ll}
     \toprule
     Symbol & Variable Name  \\ 
      \midrule
      $\Delta \Theta^{TO,U}$        & Ultimate Top Oil Temperature Rise [$^\circ$C]\\
      $\Delta \Theta^{TO,R}$        &  Rated Top Oil Temperature Rise [$^\circ$C]\\
      $L^{PU}$                      &  Load [per unit]\\
      $R$                           &  Ratio of Rated Load Loss to No-Load Loss [unitless]\\
      $n$                           &  Empirical Constant \\
      $\tau^{TO}$                   &  Oil Time Constant [s]\\
      $\tau^{TO,R}$                 &  Oil Time Constant at Rated Load [s]\\
      $m$                           &  Empirical Constant\\
      $\Delta \Theta^{TO}$          &  Top Oil Temperature Rise [$^\circ$C]\\
      $\Delta \Theta^{H}$           &  Hottest Spot Temperature Rise [$^\circ$C]\\
      $\Delta \Theta^{H,R}$         &  Hottest Spot Temperature Rise at Rated Load [$^\circ$C]\\
      $\Theta^H$                    & Hottest Spot Temperature [$^\circ$C]\\
      $\Theta^A$                    & Ambient Temperature [$^\circ$C]\\
      $F^\textrm{AA}$               & Rate of Accelerated Aging [unitless]\\
      $LT_{\textrm{normal}}$        & Normal Transformer Lifetime [h]\\
      $F^\textrm{EQA}$              & Equivalent Aging Factor [unitless]\\
      $\% LOL$                      & Percent Loss of Life\\ 

      \bottomrule
\end{tabular} 
\vspace{.5em}
\caption{Transformer Aging Variables}
\label{tab:transformer_aging_variables}
\end{table}

A transformer's lifetime is correlated with its hottest spot temperature (HST), $\Theta^H$, or the point in the coils with the highest temperature. Exceeding a threshold temperature for long periods of time causes increased aging, so the nominal power limit corresponds to a hottest spot temperature that can be held without abnormal detrimental aging effects. We measure the transformer's aging via percent loss of life over a certain elapsed time period. Calculating the percent loss of life requires tracking the hottest spot temperature evolution over time using these equations:

\begin{align}
    \Delta \Theta ^ {TO,U}_{t+1} &= \Delta \Theta^{TO,R} \left( \frac{\left(L^{PU}_{t+1}\right)^2 R+1}{R+1} \right)^n \label{delta_TOU} \\ 
    \tau^{TO}_{t+1} &= \tau^{TO,R} \frac{\left(\frac{ \Delta \Theta ^ {TO,U}_{t+1}}{ \Delta \Theta ^ {TO,R}}\right)-\left(\frac{ \Delta \Theta ^ {TO}_{t}}{ \Delta \Theta ^ {TO,R}}\right)}{\left(\frac{ \Delta \Theta ^ {TO,U}_{t+1}}{ \Delta \Theta ^ {TO,R}}\right)^\frac{1}{n}-\left(\frac{ \Delta \Theta ^ {TO}_{t}}{ \Delta \Theta ^ {TO,R}}\right)^\frac{1}{n}} \label{tau_TO}\\
    \Delta \Theta ^ {TO}_{t+1} &= \left( \Delta \Theta ^ {TO,U}_{t+1} - \Delta \Theta ^ {TO}_t \right) \left( 1-\exp\left(\frac{-t}{\tau^{TO}_{t+1}}\right) \right) \nonumber \\
   & + \Delta \Theta ^{TO}_t  \label{top_oil}\\
   \Delta \Theta^H_{t+1} &=  \Delta \Theta ^{H,R} \left(L^{PU}_{t+1}\right) ^{2m}  \label{delta_H} \\
     \Theta^H_t &= \Theta^A_t + \Delta \Theta^{TO}_t + \Delta \Theta_t ^ H \label{HST}
\end{align}

Equations~\eqref{delta_TOU},~\eqref{tau_TO}, and~\eqref{top_oil} govern the change in top oil temperature due to transformer loading, and Equation~\eqref{delta_H} calculates the hottest spot temperature rise due to loading. Equation~\eqref{HST} yields the hottest spot temperature at each time step by summing the ambient temperature, top oil rise, and hottest spot temperature rise.

From the hottest spot temperature, we then calculate the rate of accelerated aging, equivalent aging factor, and the percent loss of life:

\begin{align}
     F^\textrm{AA}_t &=\exp\left(\frac{15000}{383} - \frac{15000}{\Theta^H_t + 273}\right) \label{faa} \\
    F^\textrm{EQA} &= \frac{\sum_{t=1}^{T} \Delta t F^{aa}_t }{ \sum_{t=1}^{T} \Delta t }\label{feqa} \\
    \% LOL &= \frac{100 F^\textrm{EQA} T}{LT_\textrm{normal}} \label{LOL}
\end{align}
The rate of accelerated aging $F^\textrm{AA}$ is a function of hottest spot temperature (Equation~\eqref{faa}). Integrating $F^\textrm{AA}$
over the simulation period and dividing by total simulation time yields the the equivalent aging factor $F^\textrm{EQA}$ shown in Equation~\eqref{feqa}. Lastly, the percent loss of life shown in Equation~\eqref{LOL} relates the equivalent aging factor with the rated lifetime, $LT_\textrm{normal}$.

  \section{Case Study}
  \label{case_study}


To simulate each optimization scheme, we analyze a dataset with 48
 unique homes in San Jose, California in 2018. We use simulated home load data from NREL ResStock~\cite{wilsonEndUseLoadProfiles2022}; each home has a solar PV 
array and home loads. We generate Level 2 home EV charging data to assign to each home from the NREL OCHRE residential energy model~\cite{blonsky2021ochre}.
 We assume a perfect power factor and that the battery is able to execute commands exactly 
(i.e. there is no error or bias in the battery's setpoint and sensor readings).

To calculate the cost of electricity for consumers, we construct a cost vector $\textbf{C}$ based on PG\&E's EV2-A TOU 
rate plan~\cite{pgeResidentialRatePlan2024}. The total simulation time horizon is four weeks and we separately simulate four weeks in January and July 2018 beginning on a Monday (January 8 and July 2, respectively). The MPC horizon is one day. These and other simulation parameter values are 
listed in Table~\ref{tab:params}. In this table, the battery capacity and power limit parameters indicate those of each 
individual battery, not the aggregate. The transformer penalty weight, $\lambda$, is set to $\lambda=100$ to balance the system electricity costs with the transformer penalty value. The final electricity costs reported do not include this penalty value.

For each simulation month, we run 50 trials, sampled as described in Section~\ref{violations}. This sampling yields a different number of homes in each trial for January and July. Based on this random selection of homes, we observe between
6 and 22 homes located downstream of the transformer. In particular, in January, each trial has an average of 12.6 homes with a standard deviation of 2.7 homes and in July, each trial has an average of 16.1 homes with a standard deviation of 3.4 homes.

\begin{table}[ht]
    \centering
    \begin{tabular}{l  r | l r} 
     \toprule
     \multicolumn{1}{l }{Parameter}
     &  \multicolumn{1}{r|}{Value} & 
     \multicolumn{1}{l}{Parameter}
     &  \multicolumn{1}{r}{Value}\\  
     \midrule
     $T^{\textrm{MPC}}$    & 24       &    $\Delta \Theta^{TO,R}$             & 65$^\circ$C \\
     $T$                   & 1344     &    $\tau^{TO,R} $                     & 8.738 s   \\
     $\lambda$             & 100      &    R                                  & 3.625  \\
     $\alpha$              & 0.01     &    $\Theta^A$                         & 25$^\circ$C\\
     $\Delta t$            & 0.5 h    &    $\Delta \Theta^{H,R}$              & 15$^\circ$C  \\
     $\eta$                & 0.9487   &    $LT^\textrm{normal}$               & 180000 h     \\
     $E^{\textrm{max}} $   & 13.5 kWh &    $C^\textrm{batt}$ & \$5550 \\ 
     $B_\textrm{max}^\textrm{dischg}$ & 5 kW  &  $N^\textrm{cyc}$   & 1400\\
     $B_\textrm{max}^\textrm{chg} $  & 5 kW & &  \\

     \bottomrule
\end{tabular}
\vspace{.5em}
\caption{Parameter Values}
\label{tab:params}
\end{table}

  \section{Results}
  \label{results}
  To highlight the impact of virtualization on battery control, we compare the individual, joint, hybrid, and dynamic schemes to a system without a battery. Analyzing both transformer protection and cost reduction metrics reveals differences in each scheme's performance. We also examine the effect of changing battery partition sizes in the hybrid case. Next, we analyze the sensitivity of the schemes to different pricing plans, imperfect forecasts, different seasons, varying EV penetration levels, and battery aging costs. Finally, we show experimental results from a small-scale physical testbed.

\subsection{Meter and Transformer Impacts of Battery Aggregation}
\label{sec:5.1}

\begin{figure}[ht!]

   \centering
       \includegraphics[width=\linewidth]{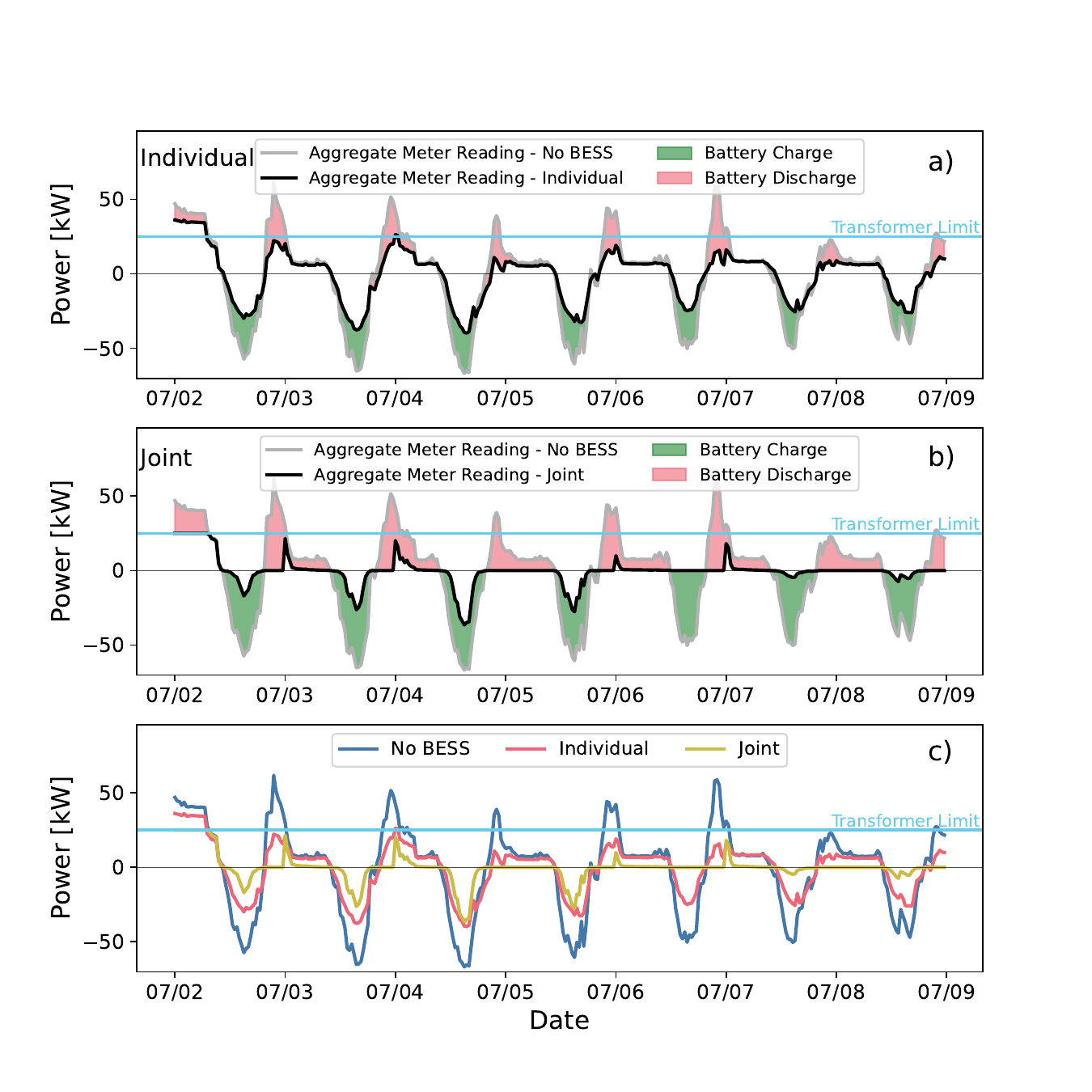}
       \caption{Aggregate meter comparison between (a) individual, (b) joint, and (c) both schemes for the first week in July 2018. The gray curve indicates the aggregate meter (sum of net meter of all homes) before adding the BESSs and the black curve indicates the meter after adding the BESSs. (c) overlays the no BESS, individual, and joint optimization, highlighting the joint scheme best utilizes available solar generation. By sharing the BESSs, the joint scheme can maintain the meter under the transformer limit for the whole week, while the individual scheme cannot.}

 \label{fig:2_meter}
 \end{figure}

While the individual scheme, in which each homeowner controls their own BESS, shows a drastic improvement over a system with no BESSs, it still does not fully utilize the full BESS potential. Figure~\ref{fig:2_meter} compares the aggregate meter values for the system in the first week of July 2018 for a single trial. On each of the plots (individual scheme in Figure~\ref{fig:2_meter}a and joint scheme in in Figure~\ref{fig:2_meter}b), the black curve shows the aggregate (summed) net meter for the collection of homes in the trial. The gray curve shows the net meter with no BESSs present in the system. The shaded areas indicate the BESS behavior that makes up the difference between the black and gray curves (green shading is charging while red shading is discharging). As shown in the green shaded areas, the BESS takes advantage of excess solar to charge. It uses this to discharge in the evening times to offset high evening loads (shaded red).

 The joint scheme maintains the aggregate meter under the transformer limit for the entire simulation period, but the individual scheme violates the limit in the first few hours of the week, shown by the black curve exceeding the blue transformer limit line. This result is highlighted in the direct comparison in Figure~\ref{fig:2_meter}b. The joint scheme utilizes the full aggregate BESS capacity to minimize excess solar generation that would otherwise be curtailed. The individual scheme runs out of battery capacity to capture this excess solar. For example, on 7/6, the joint scheme BESS uses all of the excess solar, which the individual scheme cannot do.

\begin{figure}[ht!]
  \centering
      \includegraphics[width=\linewidth]{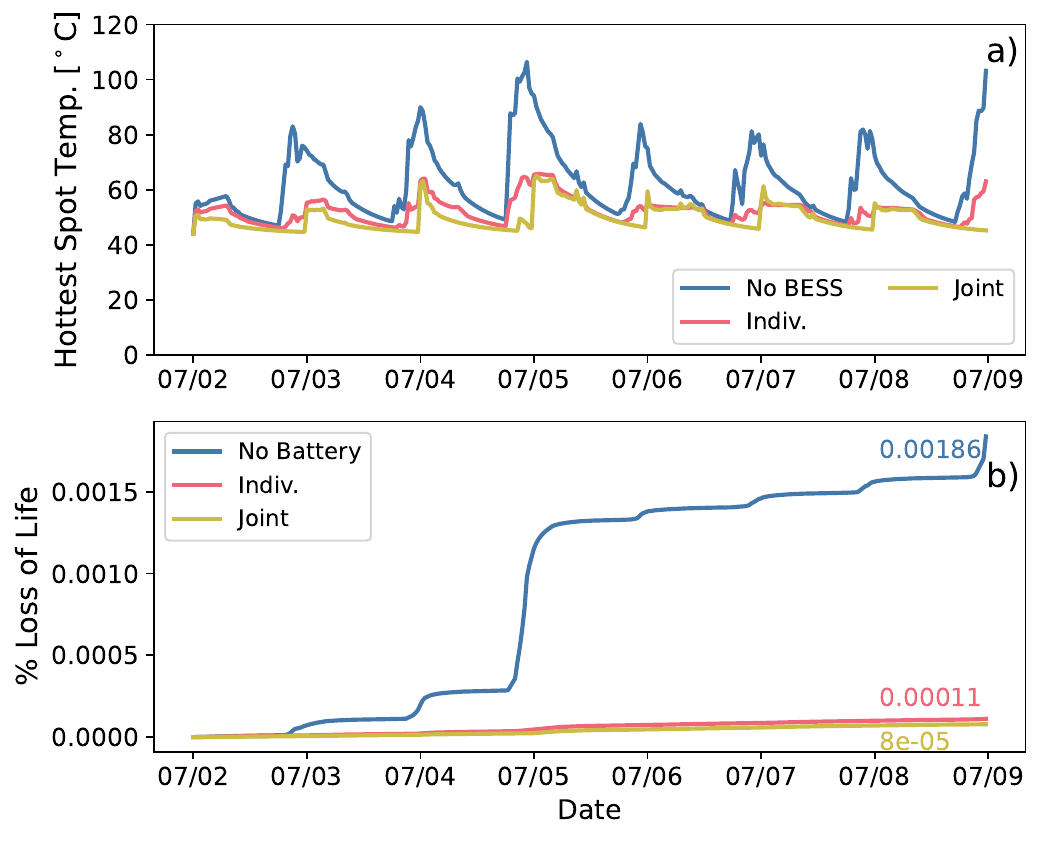}
      \caption{(a) HST with no BESS compared to individual and joint optimization. For this week of simulation in July 2018, having no BESS substantially raises HST values during the load peaks. (b) Percent loss of life for no BESS, individual, and joint optimization. The HST spikes correlate directly to larger increases in loss of life. }
\label{fig:3}
\end{figure}

Figure~\ref{fig:3} displays the HST and percent loss of life during the first week of July 2018 for the same single trial. 
While there is a cumulative low loss of life without a BESS, the HST often spikes more than 20$^\circ$C above the individual and joint schemes with a BESS, 
whose HST remains below 70$^\circ$C during this week. The individual and joint schemes have similar losses of life, underscoring that 
just adding BESSs to a system without aggregation still has a substantial effect on transformer protection. Lastly, this plot demonstrates that loss of 
life is monotonically increasing: no loading pattern can decrease loss of life.

\subsection{Transformer Loss of Life and Cost Reduction Metrics}

\begin{figure}[ht!]

  \centering
      \includegraphics[width=\linewidth]{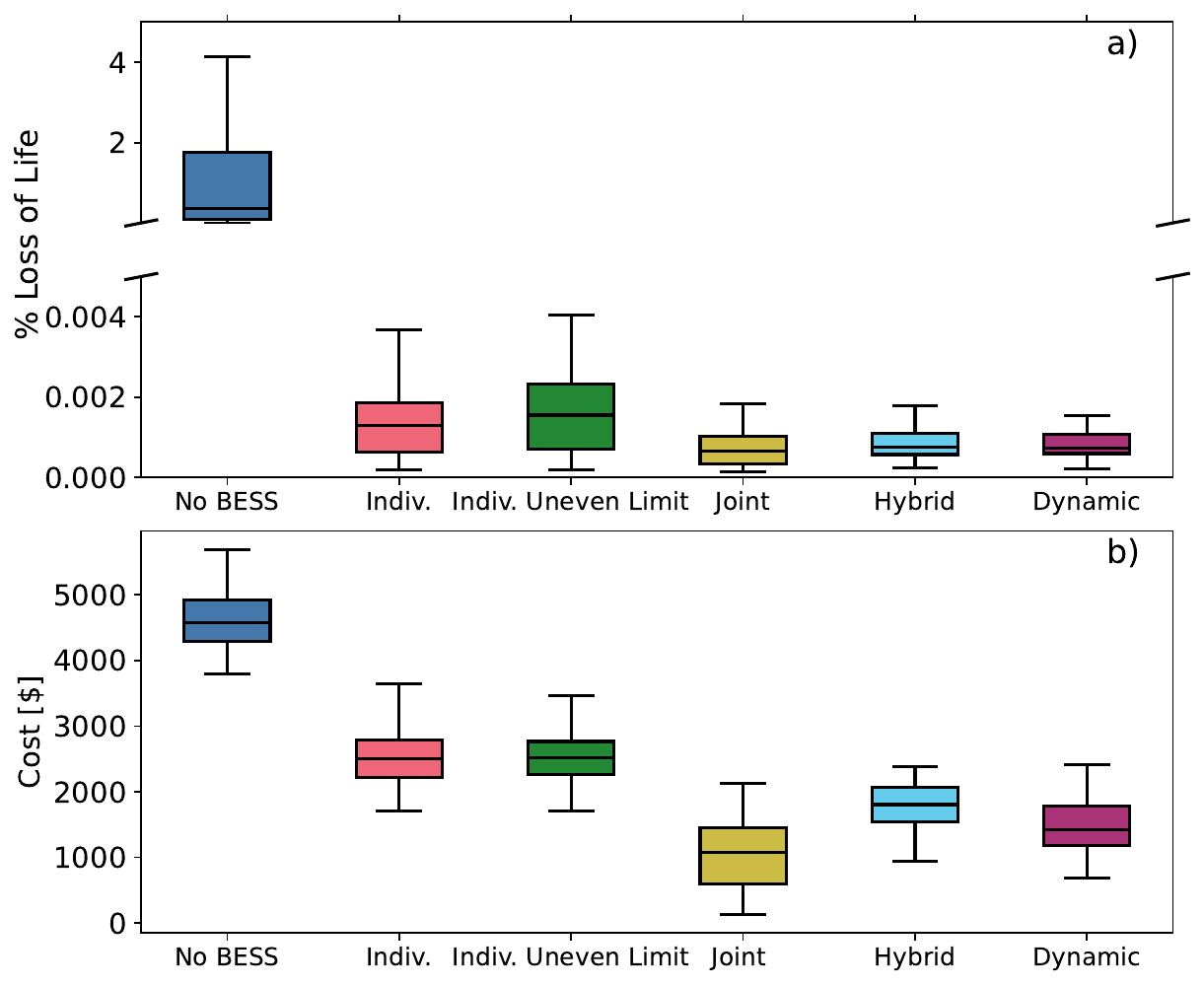}
      \caption{(a) Percent loss of life over the month of July 2018 comparing each scheme. (b) Total system electricity costs for July 2018. The boxplots indicate results from 50 trials. Including a BESS decreases transformer aging substantially, and introducing aggregation through virtualization has a strong effect on cost reduction. Joint optimization, or complete sharing, performs the best in both transformer and cost metrics}

\label{fig:4_t_c}
\end{figure}

Given similarities in transformer protection, the main difference between the individual and joint schemes lies in the total electricity costs. 
Figure~\ref{fig:4_t_c} shows the loss of life (a) and total costs (b) for each scheme for four weeks of July 2018. Each scheme is 
effective in protecting the transformer, but aggregating the batteries with the joint, hybrid, and dynamic schemes introduces cost savings of 53-56\% 
over the individual scheme.

Although the median value of loss of life with no BESS is low at 0.36\%, almost 25\% of the trials in the no BESS case exceed 2\% loss of life in July. 
Continuing at this rate of degradation, these transformers will reach their end of life in just over four years, well below the expected lifetime of 20 
years~\cite{IEEEGuideLoading2012}. For the schemes with a battery, the loss of life incurred in July is negligible, since the battery allows 
the aggregate meter to remain at or below the transformer nameplate capacity.

The differences between costs for each scheme are more distinct. The joint scheme offers a 76\% median cost decrease compared to the no BESS system cost. 
The median costs slightly rise across the joint, hybrid, and dynamic cases from \$1082, to \$1119 and \$1158, respectively. The joint case has the lowest 
total cost because it utilizes the battery most efficiently, with complete sharing.

However, the hybrid and dynamic schemes retain a portion of the battery 
that is not shared. This provides a crucial component for homeowner piece of mind: the ability to autonomously control their battery. For example, in a 
blackout scenario, a homeowner might want to leave some charge in reserve.  Or, they might want the ability to participate in lucrative demand response opportunities. 
Given similar transformer aging and costs, the hybrid and dynamic schemes are almost as effective as the joint scheme but with the additional benefit of a retained battery partition.

In the individual scheme with uneven transformer limit, each home has unique but static 
transformer limit allocation. Despite the flexibility, this scheme does not perform better than with an equal limit. Transformer loss of life is slightly worse. 
Since homes better utilize their transformer limit, the aggregate meter reads closer to the transformer nameplate limit, elevating the HST and accelerating aging. 
As costs are primarily driven by battery availability, a higher transformer allocation does not translate into cost savings in the absence of battery aggregation.

Finally, the dynamic scheme has slightly higher costs than the hybrid scheme despite having time-varying partition weights. 
This occurs because of the multi-objective problem tradeoffs between transformer aging and costs; the dynamic scheme protects 
the transformer more than the hybrid scheme does.

\subsection{Sensitivity of Hybrid Scheme to Partition Fractions}

\begin{figure}[!ht]
   \centering
       \includegraphics[width=\linewidth]{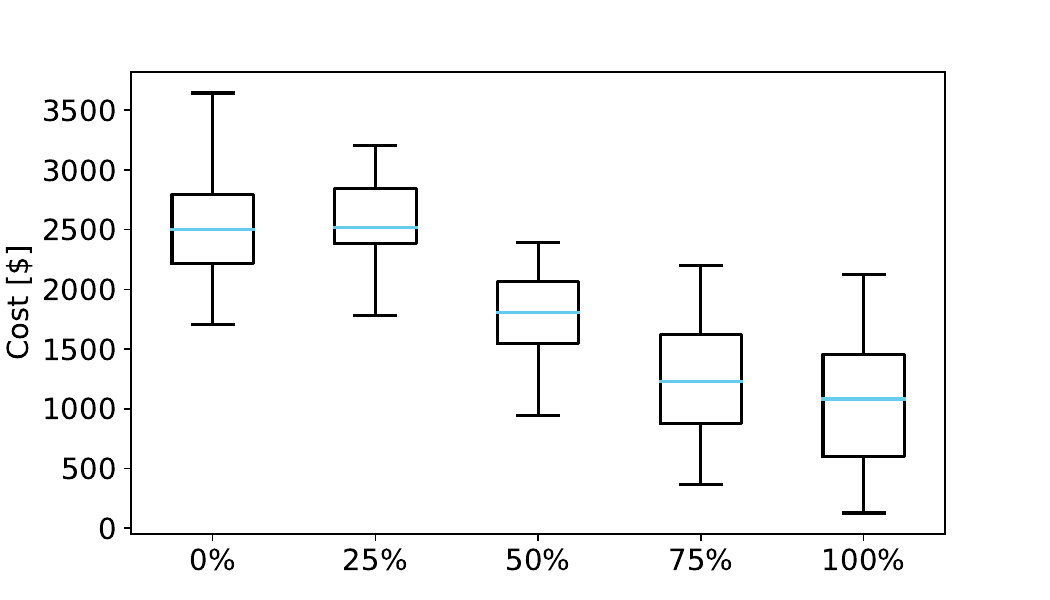}
       \caption{Sensitivity to different partition fractions for the hybrid scheme for July 2018 over 50 trials. While there is a large spread between trials, sharing only 25\% does not decrease costs, but sharing 50\% or more has substantial cost reductions. Complete (100\%) battery sharing yields the lowest costs.}

 \label{fig:5_weights_case_4} 
 \end{figure}

In Figure~\ref{fig:4_t_c}, the hybrid scheme is shown with fixed, 50\% partitions for the shared and retained battery partitions. This means each homeowner 
retains half and shares half of their battery. To examine the effect of changing this fixed partition value, Figure~\ref{fig:5_weights_case_4} varies the fraction 
of shared battery from 0\% (individual scheme) to 100\% (joint scheme). 

Sharing only 75\% of the battery yields similar total costs as the 100\% sharing. The median cost for 75\% sharing is only 14\% worse than the median for 100\% sharing. For many homeowners, this difference may be irrelevant when considering the benefits of complete control over 25\% of their battery. For example, the homeowners could recoup this 14\% cost differential by using their 25\% retained partition to participate in lucrative demand response programs. 

Understanding this critical point of 75\% sharing is important in implementing this algorithm in a real system. This results reveals that there needs to be more than 50\% sharing to achieve the cost benefits of joint optimization, but 100\% sharing is not necessary to reach these cost minimums.

\subsection{Dynamic Partitioning with Varying Electricity Pricing Plans}

When every home is subscribed to the same TOU rate structure, the dynamic scheme improves transformer loss of life but increases costs slightly compared to the hybrid scheme. 
In practice, homes in a neighborhood may be subscribed to different rate structures depending on their load usage patterns. When peak price periods are not consistent among homes, 
dynamic partitioning offers flexibility to share battery partitions when other homes need it and retain partitions for homes experiencing peak pricing.

To test this, we run a modified version of the individual, hybrid, and dynamic schemes in which each home is assigned a random price structure among the EV2-A, EV-B, 
and TOU-D structures. The individual term calculating the cost of the individual home meter (the first term in Equation~\eqref{eqn:hybrid_obj}) is subject to this 
randomly chosen price structure while the price structure for the joint term remains the EV2-A structure.

\begin{figure}[!ht]

      \centering 
         \includegraphics[width=\linewidth]{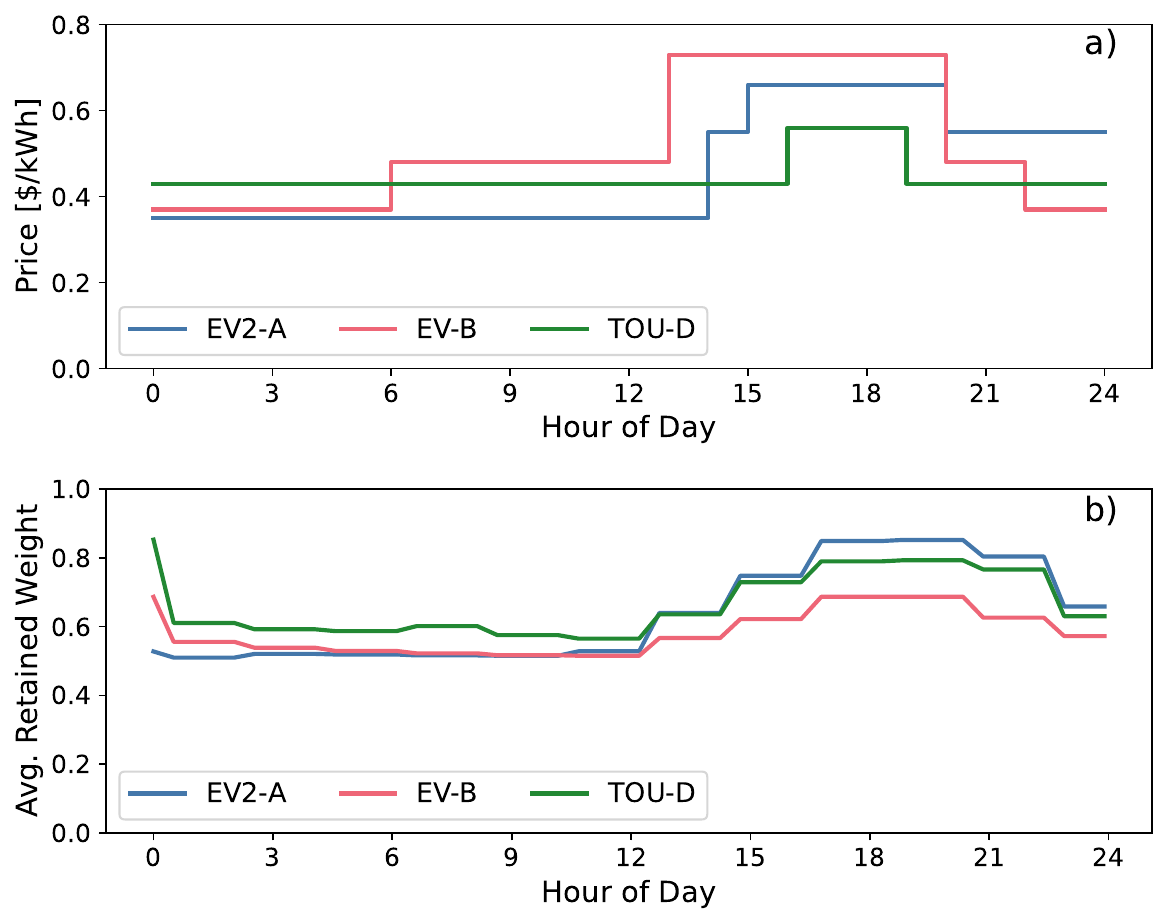}
          \caption{(a) Prices for one weekday for PG\&E EV2-A, EV-B, and TOU-D rate structures. Both the magnitude and timing of peak pricing varies between the rate structures. (b) Average retained partition weight for each timestep for the dynamic scheme. The weights change corresponding to changes in the rate structures. }
    \label{fig:6}
    \end{figure}

Figure~\ref{fig:6}a compares the three price structures, which vary by both peak price value and timing. Figure~\ref{fig:6}b plots the average retained partition weight (i.e., the partition of the battery kept by the homeowners) for the dynamic scheme, separating each group of homeowners by their pricing structure. All three groups' weights step up or down at each time when a pricing structure changes. This reflects the dynamic scheme's ability to re-adjust the partitions so the operation is cost-optimal. The retained weight is higher during peak times when the prices are highest, which is expected as homes need a greater fraction of the battery to offset home loads during peak periods.

\begin{figure}[!ht]

      \centering 
         \includegraphics[width=\linewidth]{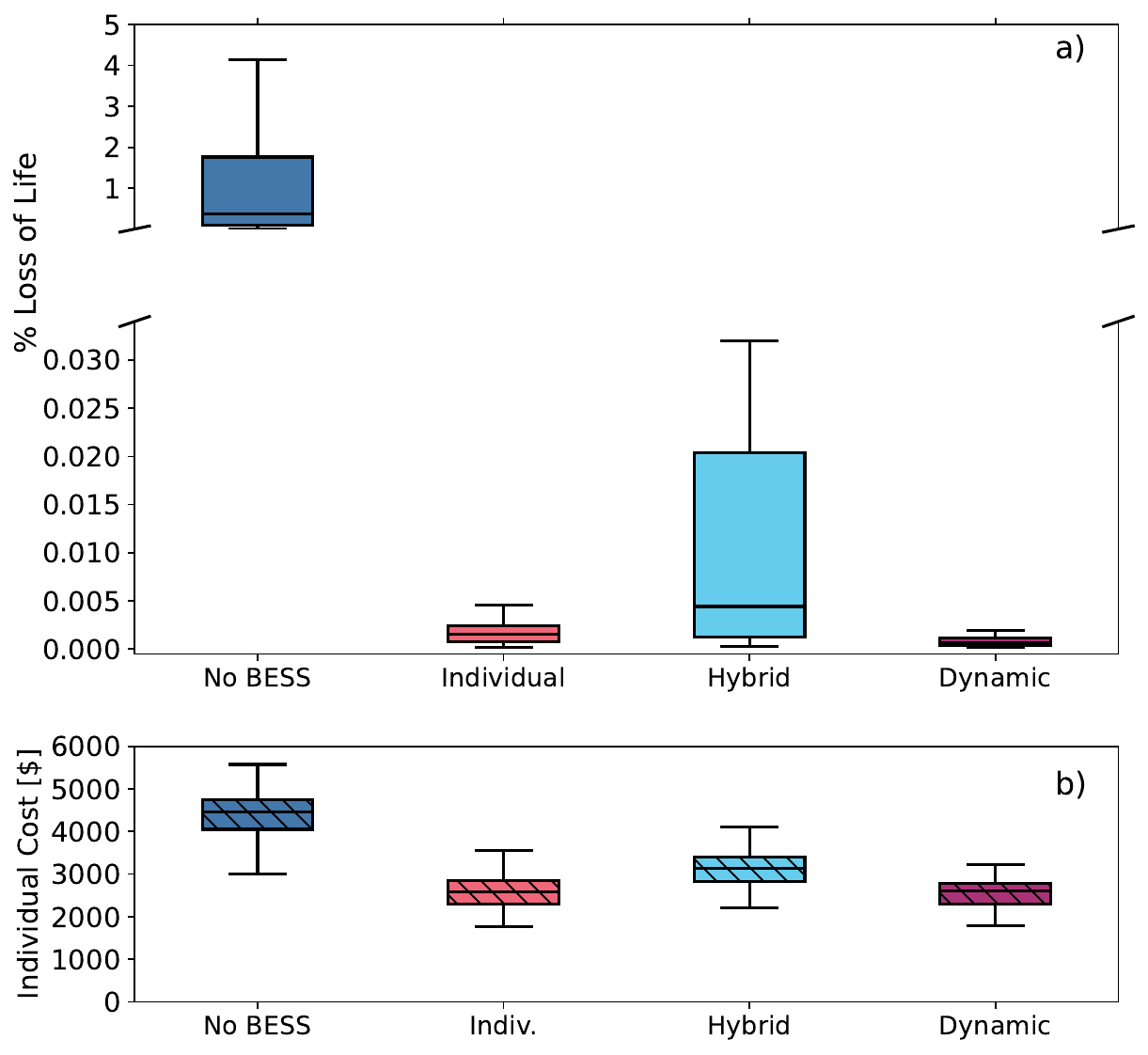}
          \caption{(a) Loss of life for no BESS, hybrid with 50\% weights, and dynamic schemes for 50 trials in July 18 with homes on varying rate structures. (b) Sum of individual meter costs for each scheme, representing the first term in the objective function for hybrid and dynamic schemes. The dynamic scheme has similar costs to the individual scheme due to its temporal flexibility. }
    \label{fig:7}
    \end{figure}

Figure~\ref{fig:7}a shows the transformer 
loss of life for each scheme compared to the no BESS scenario. The dynamic scheme yields the lowest transformer loss of life. Figure~\ref{fig:7}b
examines the individual partition costs for each scheme, indicated with hatching to differentiate from the total costs shown in Figure~\ref{fig:4_t_c}. 
For the hybrid and dynamic scheme, this is calculated by the optimal value of the first term in Equation~\ref{eqn:hybrid_obj}.

It is expected that the hybrid scheme has higher costs than the individual scheme, since only 50\% of each home's battery is represented in this 
cost calculation while the individual scheme utilizes 100\%. However, although the dynamic scheme shares only 38\% and retains 62\% of the battery on average across all homes and 
trials, it still achieves similar costs as the individual scheme. This result demonstrates the value of virtualization's temporal flexibility. 
By changing partitions over time, the dynamic scheme can retain only the minimum necessary battery fraction and share the remainder, 
leading to similar costs as the individual scheme but better transformer protection.

\clearpage 

\subsection{Sensitivity to Imperfect Forecasts}

All the simulation results above represent optimization algorithms with perfect foresight, meaning the problem is solved with knowledge of exact solar and load profiles over the simulation horizon. In practice, this is not the case as forecasts are often imperfect. MPC adjusts for this by re-solving the optimization problem at each timestep with a short forecast horizon and updating the SOC based on actual solar and load values once they occur.

\begin{figure}[ht!]

  \centering
   \includegraphics[width=\linewidth]{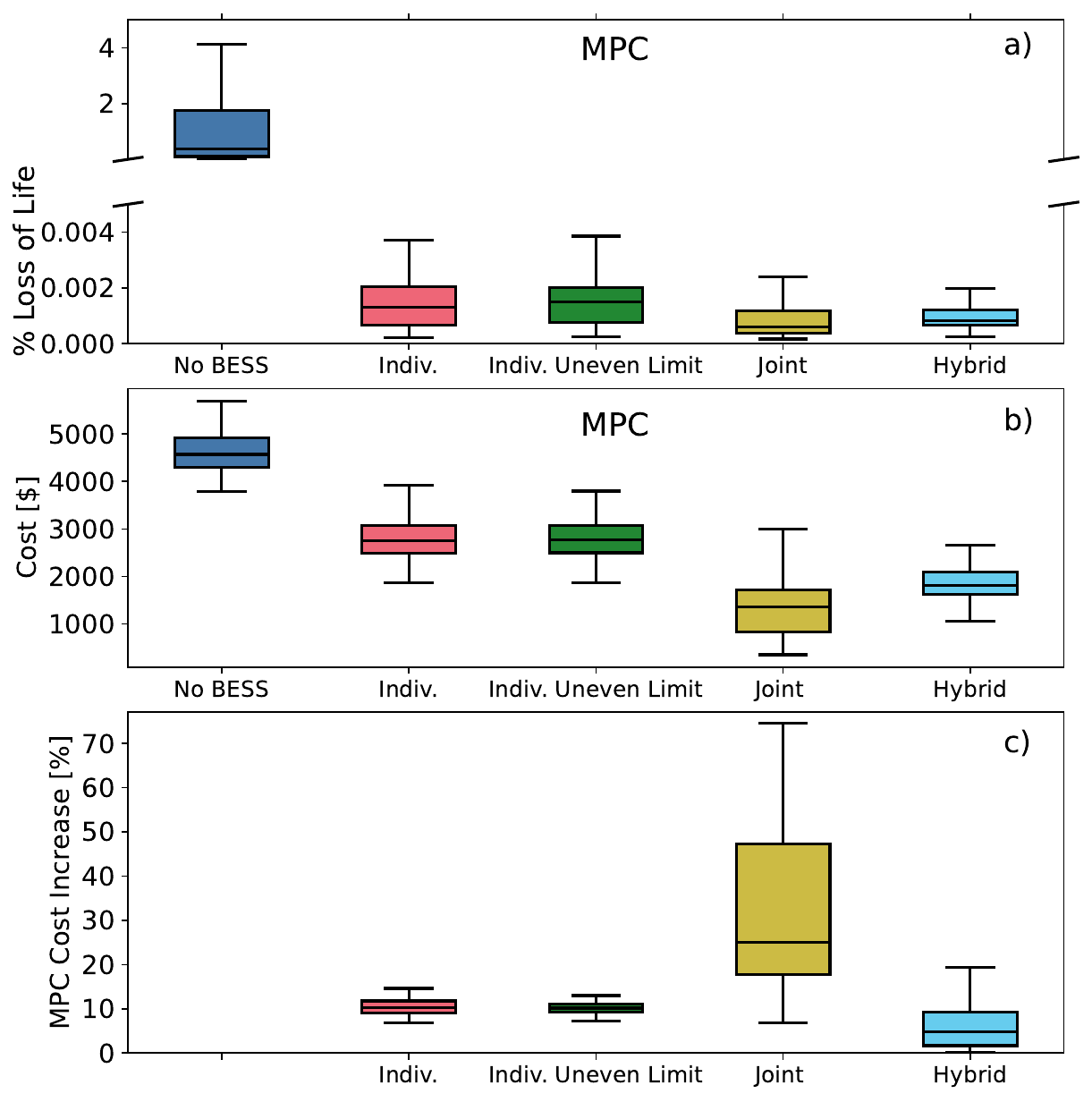}
      \caption{(a) Percent loss of life over the month of July 2018 comparing each scheme solved with MPC. Each boxplot represents results for 50 trials. (b) Total system electricity costs for July 2018 with MPC. (c) Percent cost increases for each scheme using MPC compared to perfect foresight. The boxplots indicate results of 50 trials. Using MPC has minimal impact on transformer aging and total cost, with all schemes yielding less than 1\% cost increases monthly.}

\label{fig:8_MPC}
\end{figure}

Figure~\ref{fig:8_MPC} shows the transformer aging and cost for the schemes using MPC instead of perfect foresight. We do not test MPC on the dynamic scheme as its performance depends on perfect load and solar foresight over the entire time horizon. Overall, transformer aging and costs follow the same patterns as seen with perfect foresight control. The individual scheme is not as cost-effective as the joint scheme. However, the individual scheme is still preferable to the no BESS system, even with the slightly worse performance of MPC.

Figure~\ref{fig:8_MPC}c shows the percent cost increase for each scheme using MPC compared to the cost using perfect foresight. Cost increases for individual and individual with uneven limit are around 10\%. The cost increase spread grows with the schemes that incorporate battery sharing, but the median differences are still less than 25\%. This discrepancy is because with a shared battery, an inaccurate forecast for one home may affect the entire shared battery, but with the individual scheme, the inaccurate forecast would be confined to just a single home. This phenomenon explains the larger cost increase spread for the joint and hybrid schemes. The deviations for transformer aging are also small; the greatest aging increase for July 2018 is 5\% for joint optimization. We also find that these results hold across seasons such that battery sharing with MPC is still cost-optimal compared to the No BESS case.

 The MPC cost increases correspond to a range between a \$91.96 increase for the hybrid scheme to \$269.31 for the individual scheme. Although the joint scheme has the greatest percentage median cost increase, in unnormalized dollars, the individual scheme has the greatest increase. The results show that even with a small cost increase, MPC implementation is still preferred to the No BESS case. With better forecasting, these increases can be mitigated even further. Ultimately, the cost increases only amount to between \$5.75 and \$18.5 per home compared to perfect foresight, indicating that deploying these algorithms in a real system is economically viable.

\subsection{Seasonality Effects}
\label{sec:5.6}
\begin{figure}[ht!]

  \centering
      \includegraphics[width=\linewidth]{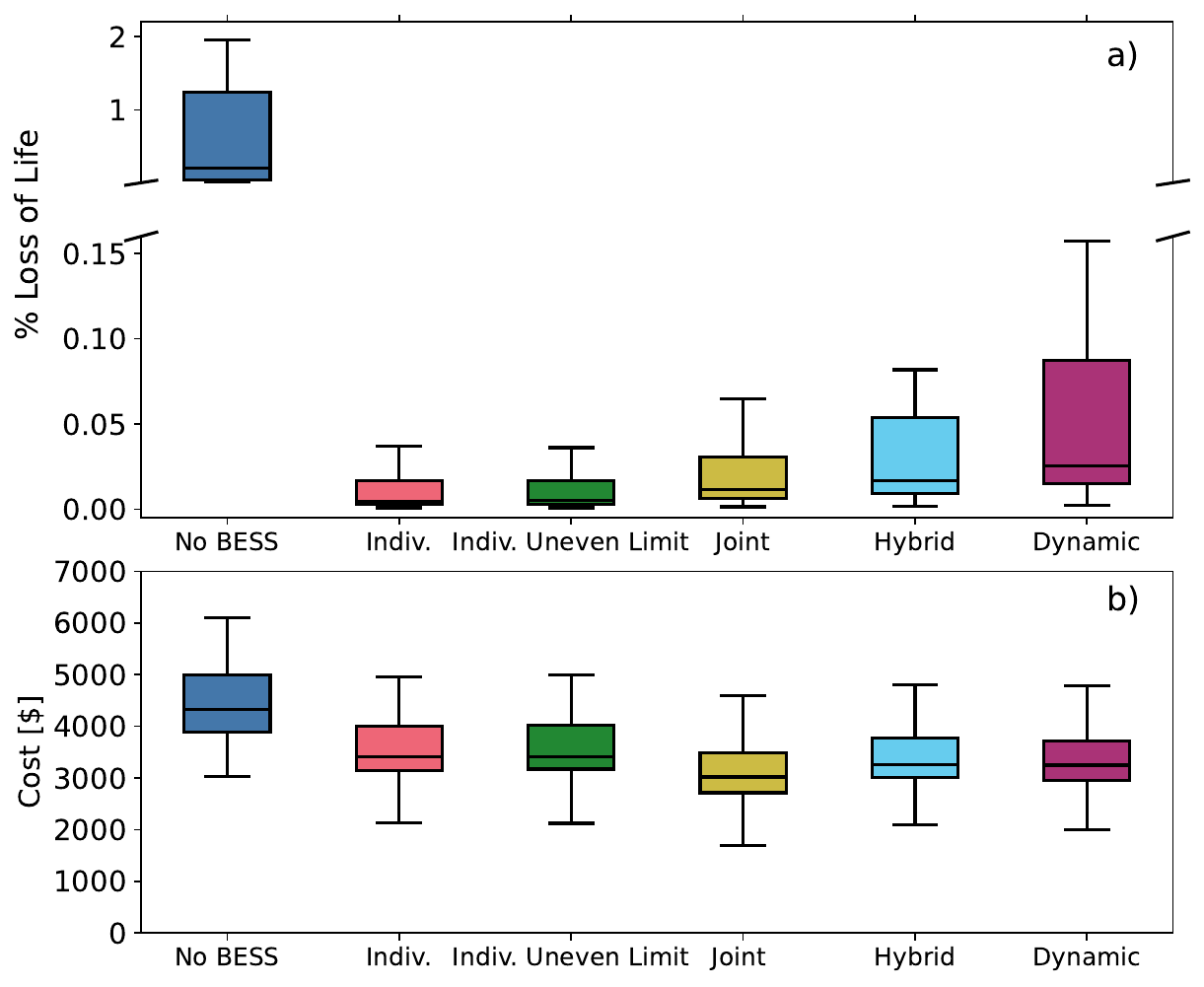}
      \caption{(a) Percent loss of life over the month of January 2018 comparing each scheme. (b) Total system electricity costs for January 2018. The boxplots indicate results of 50 trials. Due to limited solar, loss of life is worse and cost reduction is not as effective compared to July performance.}

\label{fig:9_january}
\end{figure}

The July simulation results are not necessarily representative of system behavior during the rest of the year. To ensure the optimization yields similar transformer protection and cost savings, Figure~\ref{fig:9_january} presents results for each optimization scheme for January 2018.

In January, owning a BESS is still effective, but the cost savings are not as drastic. Although transformer loss of life is an order of magnitude worse than July, the aging still remains under 0.08\% for all schemes. On the cost side, individual and joint optimization only offer a 21\% and a 30\% cost reduction, respectively, compared to no BESS. Although January loads are lower because there is no need for air conditioning, lower solar generation is the main reason for the discrepancy between January and July. Less excess solar means there are fewer opportunities for the BESSs to charge for free from solar and discharge to offset power at expensive peak times. Regardless, these results prove that BESS optimization is worthwhile across the whole calendar year, though its impact (cost reduction or transformer protection) varies over the seasons.

\clearpage

\subsection{Sensitivity to EV Penetration and Battery Aging}
\label{sec:5.7}

\begin{figure}[ht!]

      \centering
          \includegraphics[width=\linewidth]{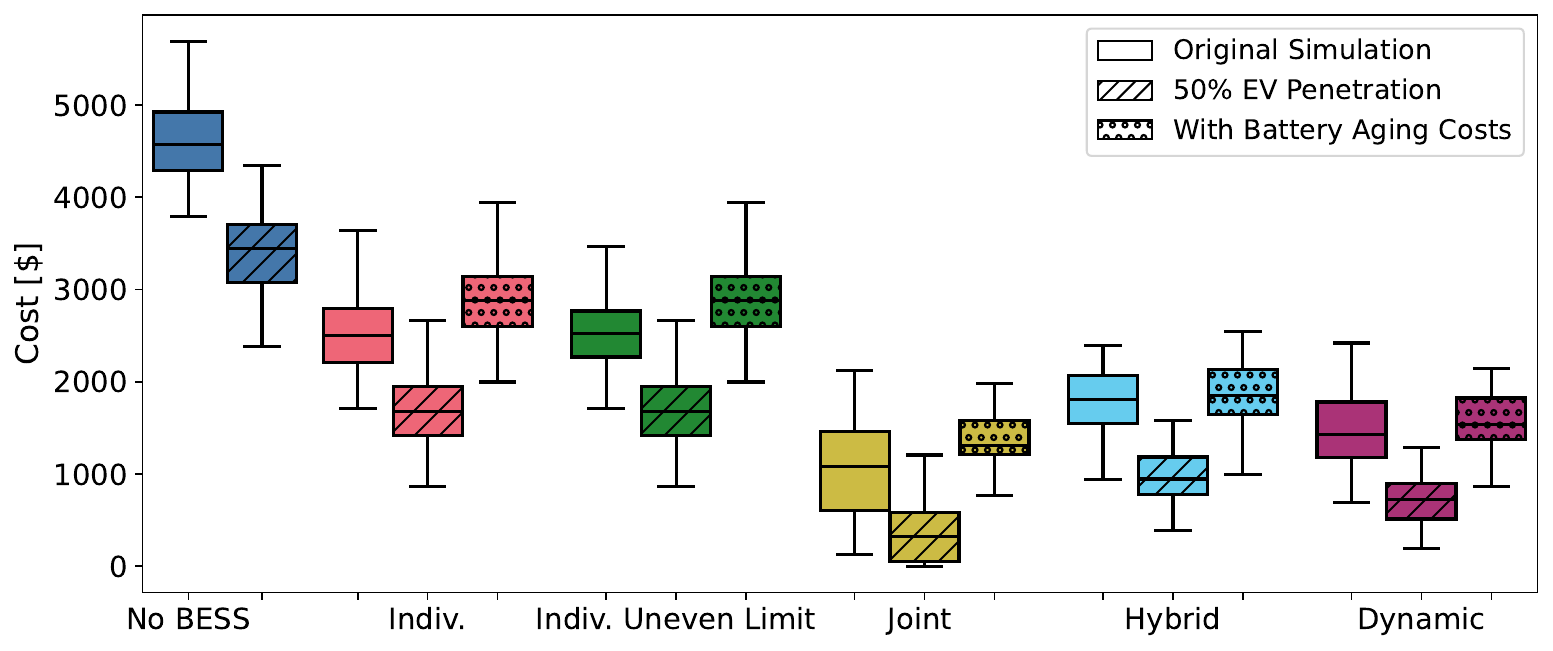}
          \caption{Comparison of July 2018 costs for each scheme between original, 50\% EV penetration, and included battery aging cost simulations. The boxplots indicate results of 50 trials. With 50\% EV penetration, the median costs are consistently lower than with 100\% EV penetration across all schemes. Adding a battery aging cost has a greater effect when the batteries are not shared (i.e., in the individual schemes).}
    
    \label{fig:10_varying_aging}
    \end{figure}

The results from Sections~\ref{sec:5.1} to~\ref{sec:5.6} assume both 100\% EV penetration (i.e., every home has one EV charger) and no economic effects from battery aging. However, in real neighborhoods, some homes may not have an EV and the cost of battery degradation may be substantial. We test the algorithm sensitivity to both EV penetration and a cost on battery aging. Figure~\ref{fig:10_varying_aging} compares the costs for each scheme between the original simulation, a simulation with 50\% EV penetration, and a simulation that includes a battery degradation cost per cycle. 

As expected, the costs are lower in each scheme with 50\% EV penetration. Joint optimization still yields the lowest cost as it best utilizes available solar generation. However, the battery aging costs do not follow a consistent pattern across all schemes. In the individual scheme, adding a cost per cycle increases the median cost by 15.1\%, but the costs only increase by 2.5\% and 8.2\% in the hybrid and dynamic schemes, respectively. This result suggests that the impact of battery aging costs is relatively smaller in these hybrid battery sharing schemes.

\clearpage

\subsection{Small-Scale Physical System Demonstration}

The results presented in Sections~\ref{sec:5.1} to~\ref{sec:5.7} are based on simulations with simulated home load, solar, and EV data. 
In real world implementation, factors such as communication delays and setpoint errors can affect 
the efficacy of the battery commands. To quantify these delays and errors, we design a small DC testbed study using a 20Ah LiFePo4 battery and a
1.6 Ah lithium-ion battery connected to a power supply that emulates the grid and an electronic load that emulates
home loads in a community. This study highlights how battery injection reduces the power pulled from the grid through
the transformer and shows battery response time to setpoint commands. 

\begin{figure}[ht!]

      \centering
          \includegraphics[width=\linewidth]{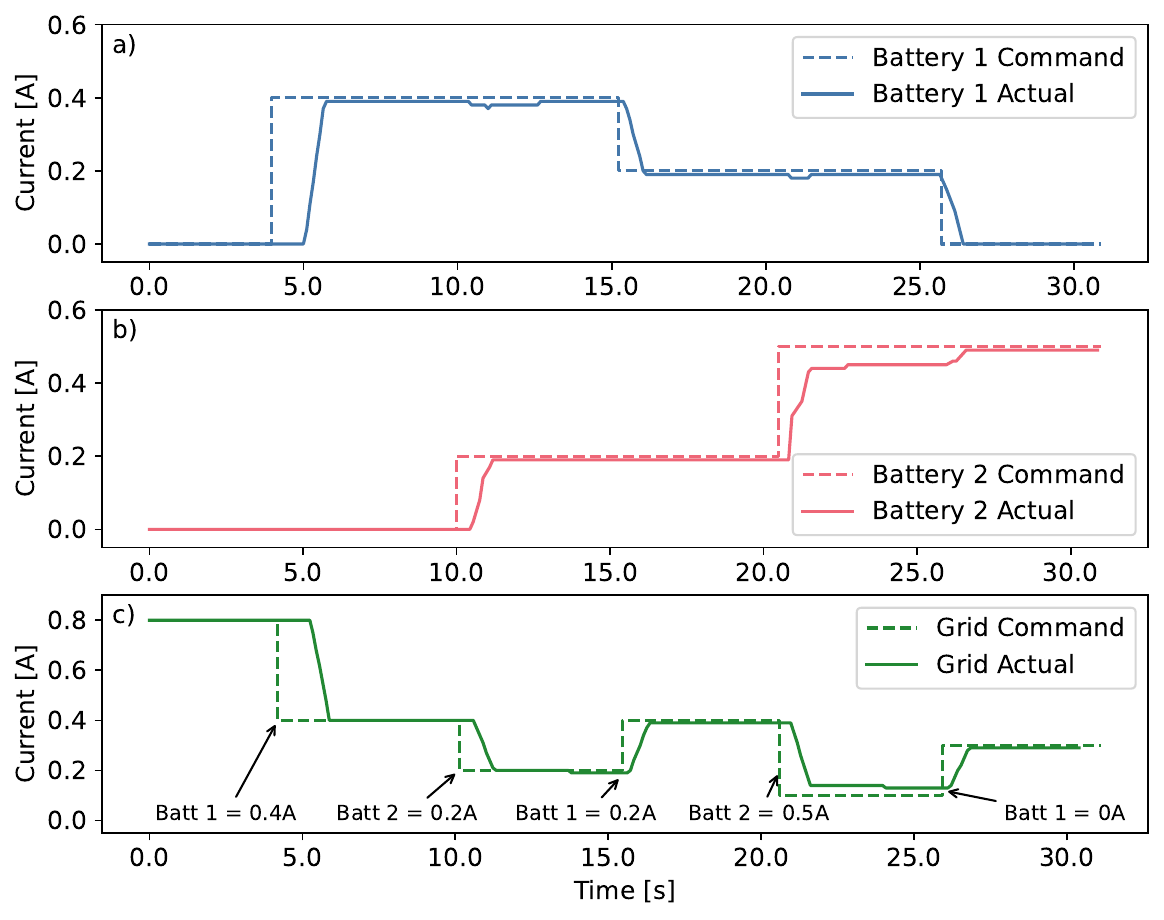}
          \caption{(a) Battery 1 command and actual current. (b) Battery 2 command and actual current. (c) Command and actual current provided by the grid, which is equal to 0.8A minus the current from batteries 1 and 2. The commands sent to batteries 1 and 2 are annotated. While there is a settling time delay, it is only 1s on average. }
    
    \label{fig:11_exp}
    \end{figure}

To illustrate setpoint errors and battery response times, we run a 30 second experiment with varying battery 
setpoint commands under a constant load. Under a constant load of 0.8A, Figure~\ref{fig:11_exp} shows the system response
to battery injection. The current being pulled through the transformer, shown in Figure~\ref{fig:11_exp}c, highlights the 
grid's response to changing battery setpoint commands. On average, the grid takes 1.00 seconds to respond to the setpoint 
commands while the batteries take 0.92 and 1.44 seconds, respectively. The setpoint errors are relatively minor for both 
batteries and the grid with RMSE values of 0.08A, 0.06A, and 0.24A, respectively. In any real world electrical system,
perturbations around the setpoint are anticipated. Overall, the communication delays are negligible under the time intervals simulated in this study. This study highlights the efficacy of the system under real world scenarios.

  \section{Discussion}
  \label{discussion}
  
The results show that all of the optimization schemes reduce transformer aging substantially over the course 
of the month-long simulation periods. Joint, hybrid, and dynamic schemes highlight the impact battery sharing has
on both reducing costs and mitigating transformer aging.

Virtualization is crucial to achieve maximum cost reduction in joint optimization.
However, with only 50\% sharing, the hybrid scheme reduces the total community cost incurred during the month of July 2018 by 57\%. 
Sharing 75\% of their batteries allows homeowners to be within 14\% of the optimal lowest cost while still maintaining control over
a modest fraction of their batteries. 

The effect of variable solar generation across seasons is substantial. Reduced 
solar generation in January leads to diminished costs savings, but the role of batteries is still prominent. 
These seasonality effects may also change in different geographical locations depending on solar and other variable 
energy generation penetration.

There are slight differences in transformer protection between optimization schemes.
The patterns vary seasonally: in the summer,  battery sharing leads to slightly better transformer protection, while in 
the winter battery sharing is slightly worse for the transformer. This phenomenon results from varying battery utilization 
among the schemes. Operating \textit{at} the transformer limit results in greater aging than operating well below the limit, even 
if the system never exceeds the limit in either case. The difference between the individual scheme and individual with 
uneven transformer limit allocation scheme also occurs because of this utilization pattern. By optimizing each home's 
limit allocation, each home is more likely to operate closer to their limit, thus pushing the aggregate meter further 
toward the nameplate limit.

Overall, complete sharing of batteries (joint optimization) is the most effective scheme for minimizing community costs 
and mitigating transformer aging. The results show a 76\% and 30\% decrease in costs compared to the no BESS case in July and January 2018, respectively.  
In the worst case of joint optimization in January 2018, 
the 0.07\% loss of life over the month would still yield a transformer lifetime of 119 years. 

While the hybrid and dynamic schemes have comparable transformer protection as pure joint optimization, they are less cost-effective. Partial sharing of the batteries allows the scheme to mitigate the worst of the peak loads to effectively reduce aging, but without complete sharing, there is still some underutilized storage accounting for the slightly higher total operational costs.

The schemes are robust to both varying EV penetration and a battery aging cost. Similar trends hold when testing sensitivity to those parameters: joint optimization is still the most cost-effective, but by leveraging virtualization, hybrid and dynamic optimization are preferable to the individual cases while still allowing homeowner autonomy.

\subsection{Implementation Considerations}

Each of the schemes provides value for transformer aging when compared to systems without batteries, but the
flexibility of the hybrid and dynamic schemes can better meet various goals and requirements of the community members. To implement these schemes in practice, there are many additional considerations including imperfect load and 
solar forecasting, cost structure, battery aging, and algorithm scalability.

 First, the MPC algorithm is susceptible to 
inaccurate solar or load forecasts, making the batteries charge at inopportune times. This can occur in two 
scenarios: when actual loads are greater than forecasted loads or when actual solar generation is less than forecasted generation. This 
inaccurate forecasting can lead to suboptimal battery behavior that violates the transformer 
limit momentarily. Ultimately, minimizing transformer aging and costs with MPC is sensitive to forecasting 
accuracy and further work will explore MPC performance with a more accurate forecast. 

Deploying the joint, hybrid, or dynamic scheme also requires a new electricity cost structure. This raises questions about cost 
fairness, battery partition sizes, and responsibility for system upgrades, operation, and maintenance costs. The utility 
must decide how to bill shared grid imports and exports and divide costs fairly among homeowners. To implement a new cost structure, each home 
must have its own meter and be able to accurately measure its load, solar, and battery output. A central aggregator could 
serve as a global controller and communicate with the utilities and homeowners about pricing, but the exact 
cost structure is dependent on existing local distribution grid infrastructure.

Battery aging is another concern in real world implementation. In each scheme, we find that the battery generally completes one full discharge 
and charge cycle per 24 hours. This usage in the individual scheme is covered under Tesla's warranty policy. Tesla 
provides a ten year warranty with unlimited battery cycling for solar backup and load shifting or a 1,400 charge and discharge 
cycle limit for other uses~\cite{teslaTESLAPOWERWALLLIMITED2025}. If homeowners use their battery partition for other programs such as grid 
frequency response or ancillary services, then battery usage should be restricted to using 140 full cycles per year, or 38\% of the 
battery capacity per day. 

When adding a cost per cycle to the simulation, the total electricity cost increases for each scheme, from 2.5\% to 21.2\% depending on the scheme. 
This reflects that when adding a cost per cycle, it is financially advantageous for the system to complete fewer cycles to prolong the battery lifetime, even if doing so increases electricity costs. 
However, this cost increase associated with battery degradation still does not exceed the costs for the system with no batteries at all, indicating that batteries are still economically preferred.

While we test the algorithm on a small, 25 kVA residential transformer, the optimization schemes can be applied to systems with 
larger transformers, which may add more batteries and more decision variables in the optimization problem. Additionally, solving 
for a longer time horizon or a smaller time granularity increases the number of decision variables. Since the optimization problem 
is convex, it is scalable to a larger problem and does not pose time constraints when solving at fast frequency with MPC. 
Virtualization software can also be scaled to large systems; in a simulation of 600 virtual batteries, latency from the virtualization software accounts for only half of the total system latency~\cite{martin2022software}. Incorporating a transformer penalty that 
only relies on the transformer nameplate 
limit and not on device-specific parameters also facilitates deployment in a variety of physical systems.

In real world implementation, virtualization is imperative to distribute battery charge and discharge 
instructions to each constituent physical battery. Virtualization serves multiple roles in this distribution process, such as allowing 
the system to re-calibrate if one battery drops offline and enabling the system to change battery partition weights frequently, 
as in the dynamic partitioning scheme. Without virtualization, homeowners could either control an individual battery and risk 
underutilizing an expensive asset, or they could buy in to a centralized, shared battery with no individual autonomy. 
Virtualization makes this combination possible while ensuring the system is robust in the face of system outages, additions, or changes.

Each of these implementation challenges are critical for home and EV owners, policymakers, utilities, and aggregators to consider. While utilities and aggregators may be concerned with the scalability and flexibility of the algorithm to deploy it in heterogeneous distribution systems, homeowners may be more focused on their direct costs of buying a battery and buying into a shared battery system. On the other hand, policymakers need to understand implementation challenges and associated costs to help design rebates or incentive programs that aim to ensure equitable deployment. All of these stakeholders must collaborate to fairly and successfully deploy this shared battery system.

\subsection{Limitations}

Battery sharing algorithms have limitations. 
This work does not focus on cost allocation fairness, forecasting accuracy and prediction models, and battery control execution.

The question of how to fairly allocate the benefits of battery sharing still remains. This paper only provides an assessment of total system cost and does not report a breakdown of system costs among homeowners. In an aggregation scheme, splitting the cost savings equally among all homeowners would not be fair to homeowners who are already able to manage their loads individually. Two other potential options are scaling the savings based on a home's peak power or total energy throughput. Future work will explore the differences in cost allocation schemes, including which type of homeowner benefits from each scheme.

Our simulation also has the limitation of a simplified perfect foresight prediction model for EV charging. In reality, EV drivers may not know their exact charging needs one day in advance. A deviation from this prediction has the potential to exacerbate transformer aging with a high load, especially if multiple EVs in a neighborhood charge simultaneously. If hardware and computational capabilities allow, a short MPC timestep (i.e., quickly correcting forecasting errors) could rectify this problem. Alternatively, a smart charger or other peak load management system could serve as a backup to curtail load in case of incorrect EV demand forecasting.

Lastly, we only examine the effect of long latencies or battery setpoint control errors in a small DC testbed. In a grid-scale system, the residential batteries may not be able to be perfectly controlled. While the MPC corrects for setpoint errors, long latencies could reduce the algorithm's performance if they are beyond the seconds scale as observed in the DC testbed. To mitigate these effects, engineers should perform extensive testing with the algorithm on physical systems to understand any latencies across different battery management systems.

  \section{Conclusion}
  \label{conclusion}


This work presents a comparison between different battery sharing optimization schemes enabled by battery virtualization. The schemes aim to reduce consumer electricity bills and mitigate residential transformer aging. We find that sharing the batteries 
results in both lower costs and similar transformer protection than a scheme where owners optimize their batteries individually. While the joint scheme offers the best utilization of the batteries, the hybrid and dynamic schemes, or combinations of the individual and joint schemes, yield similar costs and transformer protection while still allowing homeowners to keep part of the battery for themselves. By employing virtualization, the dynamic scheme improves individual costs further by changing the retained and shared partition to maximize utilization. Lastly, the proposed model predictive controller allows for real time implementation that is robust to forecast error.

Future work will explore higher fidelity forecasting methods, simulations on different configurations of batteries and transformers, and experiments on physical battery systems in homes. To test and implement our algorithms in a neighborhood, a central controller or aggregator must also decide how to fairly allocate cost savings. Both fair pricing and best practices for collaboration between aggregators, utilities, and homeowners, are other areas of future research. 

Ultimately, residential battery virtualization empowers homeowners to optimize their batteries to protect shared neighborhood transformers and reduce costs for the entire community. As EV sales continue to grow and more homeowners install Level 2 EV chargers, residential distribution transformer overload will become a more pressing problem. Our BESS optimization schemes offer the flexibility for homeowners to install chargers 
without compromising neighborhood transformer lifetimes.

\section*{Acknowledgments}
This work is supported by 
the U.S. Department of Energy, Office of Electricity under Award Number DEOE0000919 and the 
Building Technologies Office IBUILD Graduate Research Fellowship. Some of the computing for this project was performed on the Sherlock computing cluster, supported by Stanford University and the Stanford Research Computing Center.

\section*{Contributions}
S.M.: 
Conceptualization, Formal analysis, Investigation, Methodology
Software, Visualization, Writing – original draft, Writing – review \& editing; O.N.: Conceptualization,	Data curation,	Formal analysis,	Investigation, Methodology, Software, Writing – original draft, Writing – review \& editing; P.L.: Conceptualization, Funding acquisition, Project administration, Supervision, Writing – review and editing; R.R.: Conceptualization, Funding acquisition, Project administration, Supervision, Writing – review and editing

\bibliographystyle{elsarticle-num}
\bibliography{references_new.bib}

\begin{thebibliography}{10}
\expandafter\ifx\csname url\endcsname\relax
  \def\url#1{\texttt{#1}}\fi
\expandafter\ifx\csname urlprefix\endcsname\relax\def\urlprefix{URL }\fi
\expandafter\ifx\csname href\endcsname\relax
  \def\href#1#2{#2} \def\path#1{#1}\fi

\bibitem{hilshey2013estimating}
A.~D. Hilshey, P.~D.~H. Hines, P.~Rezaei, J.~R. Dowds, Estimating the
  {{Impact}} of {{Electric Vehicle Smart Charging}} on {{Distribution
  Transformer Aging}}, IEEE Transactions on Smart Grid 4~(2) (2013) 905--913.
\newblock \href {https://doi.org/10.1109/TSG.2012.2217385}
  {\path{doi:10.1109/TSG.2012.2217385}}.

\bibitem{muratori2018impact}
M.~Muratori, Impact of uncoordinated plug-in electric vehicle charging on
  residential power demand, Nature Energy 3~(3) (2018) 193--201.
\newblock \href {https://doi.org/10.1038/s41560-017-0074-z}
  {\path{doi:10.1038/s41560-017-0074-z}}.

\bibitem{gong2012study}
Q.~Gong, S.~{Midlam-Mohler}, V.~Marano, G.~Rizzoni, Study of {{PEV Charging}}
  on {{Residential Distribution Transformer Life}}, IEEE Transactions on Smart
  Grid 3~(1) (2012) 404--412.
\newblock \href {https://doi.org/10.1109/TSG.2011.2163650}
  {\path{doi:10.1109/TSG.2011.2163650}}.

\bibitem{soleimani2020economic}
M.~Soleimani, M.~Kezunovic, Economic {{Analysis}} of {{Transformer Loss}} of
  {{Life Mitigation Using Energy Storage}} and {{PV Generation}}, in: 2020
  {{IEEE}}/{{PES Transmission}} and {{Distribution Conference}} and
  {{Exposition}} ({{T}}\&{{D}}), 2020, pp. 1--5.
\newblock \href {https://doi.org/10.1109/TD39804.2020.9299895}
  {\path{doi:10.1109/TD39804.2020.9299895}}.

\bibitem{sarker2017cooptimization}
M.~R. Sarker, D.~J. Olsen, M.~A. {Ortega-Vazquez}, Co-{{Optimization}} of
  {{Distribution Transformer Aging}} and {{Energy Arbitrage Using Electric
  Vehicles}}, IEEE Transactions on Smart Grid 8~(6) (2017) 2712--2722.
\newblock \href {https://doi.org/10.1109/TSG.2016.2535354}
  {\path{doi:10.1109/TSG.2016.2535354}}.

\bibitem{kaufmanRunningLowMachines2023}
A.~Kaufman, The {{U}}.{{S}}. {{Is Running Low On The Machines Needed To Avoid
  Blackouts}},
  https://www.huffpost.com/entry/transformer-shortage\_n\_64645004e4b0005c6055c541
  (May 2023).

\bibitem{li2024impact}
Y.~Li, A.~Jenn, Impact of electric vehicle charging demand on power
  distribution grid congestion, Proceedings of the National Academy of Sciences
  121~(18) (2024) e2317599121.
\newblock \href {https://doi.org/10.1073/pnas.2317599121}
  {\path{doi:10.1073/pnas.2317599121}}.

\bibitem{powell2020controlled}
S.~Powell, E.~C. Kara, R.~Sevlian, G.~V. Cezar, S.~Kiliccote, R.~Rajagopal,
  Controlled workplace charging of electric vehicles: {{The}} impact of rate
  schedules on transformer aging, Applied Energy 276 (2020) 115352.
\newblock \href {https://doi.org/10.1016/j.apenergy.2020.115352}
  {\path{doi:10.1016/j.apenergy.2020.115352}}.

\bibitem{li0222ev}
S.~Li, W.~Hu, D.~Cao, Z.~Zhang, Q.~Huang, Z.~Chen, F.~Blaabjerg, {{EV Charging
  Strategy Considering Transformer Lifetime}} via {{Evolutionary Curriculum
  Learning-Based Multiagent Deep Reinforcement Learning}}, IEEE Transactions on
  Smart Grid 13~(4) (2022) 2774--2787.
\newblock \href {https://doi.org/10.1109/TSG.2022.3167021}
  {\path{doi:10.1109/TSG.2022.3167021}}.

\bibitem{affonso2019technical}
C.~d.~M. Affonso, M.~Kezunovic, Technical and {{Economic Impact}} of {{PV-BESS
  Charging Station}} on {{Transformer Life}}: {{A Case Study}}, IEEE
  Transactions on Smart Grid 10~(4) (2019) 4683--4692.
\newblock \href {https://doi.org/10.1109/TSG.2018.2866938}
  {\path{doi:10.1109/TSG.2018.2866938}}.

\bibitem{visakh2022controlled}
A.~Visakh, B.~Thomas, M.~P. Selvan, Controlled {{Charging}} of {{Electric
  Vehicles}} to {{Reduce}} the {{Aging}} of {{Distribution Transformers}}, in:
  2022 22nd {{National Power Systems Conference}} ({{NPSC}}), 2022, pp. 24--29.
\newblock \href {https://doi.org/10.1109/NPSC57038.2022.10069054}
  {\path{doi:10.1109/NPSC57038.2022.10069054}}.

\bibitem{rossi2025smart}
F.~Rossi, C.~{Diaz-Londono}, Y.~Li, C.~Zou, G.~Gruosso, Smart {{Electric
  Vehicle Charging Algorithm}} to {{Reduce}} the {{Impact}} on {{Power Grids}}:
  {{A Reinforcement Learning Based Methodology}}, IEEE Open Journal of
  Vehicular Technology 6 (2025) 1072--1084.
\newblock \href {https://doi.org/10.1109/OJVT.2025.3559237}
  {\path{doi:10.1109/OJVT.2025.3559237}}.

\bibitem{sarmokadam2025power}
S.~Sarmokadam, M.~Suresh, R.~Mathew, Power flow control strategy for prosumer
  based {{EV}} charging scheme to minimize charging impact on distribution
  network, Energy Reports 13 (2025) 3794--3809.
\newblock \href {https://doi.org/10.1016/j.egyr.2025.03.032}
  {\path{doi:10.1016/j.egyr.2025.03.032}}.

\bibitem{latinopoulos2017response}
C.~Latinopoulos, A.~Sivakumar, J.~Polak, Response of electric vehicle drivers
  to dynamic pricing of parking and charging services: {{Risky}} choice in
  early reservations, Transportation Research Part C: Emerging Technologies 80
  (2017) 175--189.
\newblock \href {https://doi.org/10.1016/j.trc.2017.04.008}
  {\path{doi:10.1016/j.trc.2017.04.008}}.

\bibitem{navidi2023coordinating}
T.~Navidi, A.~El~Gamal, R.~Rajagopal, Coordinating distributed energy resources
  for reliability can significantly reduce future distribution grid upgrades
  and peak load, Joule 7~(8) (2023) 1769--1792.
\newblock \href {https://doi.org/10.1016/j.joule.2023.06.015}
  {\path{doi:10.1016/j.joule.2023.06.015}}.

\bibitem{endara2025design}
I.~D. Endara, D.~M. Macas, M.~P. Valverde, J.~M. Garc{\'i}a, Design of {{Load
  Management Methods Using Distributed Storage}} and {{Their Impact}} on {{The
  Aging Level}} of {{Distribution Transformers}}, in: Z.~Peng (Ed.),
  Technological {{Advancements}} and {{Future Directions}} in {{Green Energy}}
  : {{Selected Papers}} from {{ICGET}} 2024, Springer Nature Switzerland, Cham,
  2025, pp. 19--31.

\bibitem{datta2020smart}
U.~Datta, A.~Kalam, J.~Shi, Smart control of {{BESS}} in {{PV}} integrated
  {{EV}} charging station for reducing transformer overloading and providing
  battery-to-grid service, Journal of Energy Storage 28 (2020) 101224.
\newblock \href {https://doi.org/10.1016/j.est.2020.101224}
  {\path{doi:10.1016/j.est.2020.101224}}.

\bibitem{hong2020assessment}
S.-K. Hong, S.~G. Lee, M.~Kim, Assessment and {{Mitigation}} of {{Electric
  Vehicle Charging Demand Impact}} to {{Transformer Aging}} for an {{Apartment
  Complex}}, Energies 13~(10) (2020) 2571.
\newblock \href {https://doi.org/10.3390/en13102571}
  {\path{doi:10.3390/en13102571}}.

\bibitem{li2015aggregator}
J.~Li, Z.~Wu, S.~Zhou, H.~Fu, X.-P. Zhang, Aggregator service for {{PV}} and
  battery energy storage systems of residential building, CSEE Journal of Power
  and Energy Systems 1~(4) (2015) 3--11.
\newblock \href {https://doi.org/10.17775/CSEEJPES.2015.00042}
  {\path{doi:10.17775/CSEEJPES.2015.00042}}.

\bibitem{moradiamani2023data}
A.~Moradi~Amani, S.~S. Sajjadi, W.~A. Somaweera, M.~Jalili, X.~Yu, Data-driven
  model predictive control of community batteries for voltage regulation in
  power grids subject to {{EV}} charging, Energy Reports 9 (2023) 236--244.
\newblock \href {https://doi.org/10.1016/j.egyr.2022.12.089}
  {\path{doi:10.1016/j.egyr.2022.12.089}}.

\bibitem{raghuveer2025real}
{\relax Rayaprolu}.~M. Raghuveer, B.~R. Bhalja, P.~Agarwal, Real-{{Time Energy
  Management System}} for an {{Active Distribution Network}} with {{Multiple EV
  Charging Stations Considering Transformer}}'s {{Aging}} and {{Reactive Power
  Dispatch}}, IEEE Transactions on Industry Applications (2025) 1--13\href
  {https://doi.org/10.1109/TIA.2025.3552367}
  {\path{doi:10.1109/TIA.2025.3552367}}.

\bibitem{ntube2023stochastic}
N.~Ntube, H.~Li, Stochastic multi-objective optimal sizing of battery energy
  storage system for a residential home, Journal of Energy Storage 59 (2023)
  106403.
\newblock \href {https://doi.org/10.1016/j.est.2022.106403}
  {\path{doi:10.1016/j.est.2022.106403}}.

\bibitem{hafiz2019energy}
F.~Hafiz, A.~{Rodrigo de Queiroz}, P.~Fajri, I.~Husain, Energy management and
  optimal storage sizing for a shared community: {{A}} multi-stage stochastic
  programming approach, Applied Energy 236 (2019) 42--54.
\newblock \href {https://doi.org/10.1016/j.apenergy.2018.11.080}
  {\path{doi:10.1016/j.apenergy.2018.11.080}}.

\bibitem{henni2021sharing}
S.~Henni, P.~Staudt, C.~Weinhardt, A sharing economy for residential
  communities with {{PV-coupled}} battery storage: {{Benefits}}, pricing and
  participant matching, Applied Energy 301 (2021) 117351.
\newblock \href {https://doi.org/10.1016/j.apenergy.2021.117351}
  {\path{doi:10.1016/j.apenergy.2021.117351}}.

\bibitem{lee2021autoshare}
S.~Lee, P.~Shenoy, K.~Ramamritham, D.~Irwin, {{AutoShare}}: {{Virtual}}
  community solar and storage for energy sharing, Energy Informatics 4~(1)
  (2021) 10.
\newblock \href {https://doi.org/10.1186/s42162-021-00144-w}
  {\path{doi:10.1186/s42162-021-00144-w}}.

\bibitem{khanal2023optimal}
S.~Khanal, R.~Khezri, A.~Mahmoudi, S.~Kahourzadeh, Optimal capacity of solar
  photovoltaic and battery storage for grid-tied houses based on energy
  sharing, IET Generation, Transmission \& Distribution 17~(8) (2023)
  1707--1722.
\newblock \href {https://doi.org/10.1049/gtd2.12824}
  {\path{doi:10.1049/gtd2.12824}}.

\bibitem{merrington2023optimal}
S.~Merrington, R.~Khezri, A.~Mahmoudi, Optimal sizing of grid-connected rooftop
  photovoltaic and battery energy storage for houses with electric vehicle, IET
  Smart Grid 6~(3) (2023) 297--311.
\newblock \href {https://doi.org/10.1049/stg2.12099}
  {\path{doi:10.1049/stg2.12099}}.

\bibitem{walker2021analysis}
A.~Walker, S.~Kwon, Analysis on impact of shared energy storage in residential
  community: {{Individual}} versus shared energy storage, Applied Energy 282
  (2021) 116172.
\newblock \href {https://doi.org/10.1016/j.apenergy.2020.116172}
  {\path{doi:10.1016/j.apenergy.2020.116172}}.

\bibitem{kang2023multi}
H.~Kang, S.~Jung, H.~Kim, J.~Hong, J.~Jeoung, T.~Hong, Multi-objective sizing
  and real-time scheduling of battery energy storage in energy-sharing
  community based on reinforcement learning, Renewable and Sustainable Energy
  Reviews 185 (2023) 113655.
\newblock \href {https://doi.org/10.1016/j.rser.2023.113655}
  {\path{doi:10.1016/j.rser.2023.113655}}.

\bibitem{barbour2018community}
E.~Barbour, D.~Parra, Z.~Awwad, M.~C. Gonz{\'a}lez, Community energy storage:
  {{A}} smart choice for the smart grid?, Applied Energy 212 (2018) 489--497.
\newblock \href {https://doi.org/10.1016/j.apenergy.2017.12.056}
  {\path{doi:10.1016/j.apenergy.2017.12.056}}.

\bibitem{roberts2019impact}
M.~B. Roberts, A.~Bruce, I.~MacGill, Impact of shared battery energy storage
  systems on photovoltaic self-consumption and electricity bills in apartment
  buildings, Applied Energy 245 (2019) 78--95.
\newblock \href {https://doi.org/10.1016/j.apenergy.2019.04.001}
  {\path{doi:10.1016/j.apenergy.2019.04.001}}.

\bibitem{martin2022software}
S.~Martin, N.~Mosier, O.~Nnorom, Y.~Ou, L.~Patel, O.~Triebe, G.~Cezar,
  P.~Levis, R.~Rajagopal, Software defined grid energy storage, in: Proceedings
  of the 9th {{ACM International Conference}} on {{Systems}} for
  {{Energy-Efficient Buildings}}, {{Cities}}, and {{Transportation}},
  {{BuildSys}} '22, Association for Computing Machinery, New York, NY, USA,
  2022, pp. 218--227.
\newblock \href {https://doi.org/10.1145/3563357.3564082}
  {\path{doi:10.1145/3563357.3564082}}.

\bibitem{bashir2021enabling}
N.~Bashir, T.~Guo, M.~Hajiesmaili, D.~Irwin, P.~Shenoy, R.~Sitaraman, A.~Souza,
  A.~Wierman, Enabling {{Sustainable Clouds}}: {{The Case}} for
  {{Virtualizing}} the {{Energy System}}, in: Proceedings of the {{ACM
  Symposium}} on {{Cloud Computing}}, ACM, Seattle WA USA, 2021, pp. 350--358.
\newblock \href {https://doi.org/10.1145/3472883.3487009}
  {\path{doi:10.1145/3472883.3487009}}.

\bibitem{souza2023ecovisor}
A.~Souza, N.~Bashir, J.~Murillo, W.~Hanafy, Q.~Liang, D.~Irwin, P.~Shenoy,
  Ecovisor: {{A Virtual Energy System}} for {{Carbon-Efficient Applications}},
  in: Proceedings of the 28th {{ACM International Conference}} on
  {{Architectural Support}} for {{Programming Languages}} and {{Operating
  Systems}}, {{Volume}} 2, ACM, Vancouver BC Canada, 2023, pp. 252--265.
\newblock \href {https://doi.org/10.1145/3575693.3575709}
  {\path{doi:10.1145/3575693.3575709}}.

\bibitem{kumar2018stochastic}
R.~Kumar, M.~J. Wenzel, M.~J. Ellis, M.~N. ElBsat, K.~H. Drees, V.~M. Zavala, A
  {{Stochastic Model Predictive Control Framework}} for {{Stationary Battery
  Systems}}, IEEE Transactions on Power Systems 33~(4) (2018) 4397--4406.
\newblock \href {https://doi.org/10.1109/TPWRS.2017.2789118}
  {\path{doi:10.1109/TPWRS.2017.2789118}}.

\bibitem{clarke2016economic}
W.~C. Clarke, C.~Manzie, M.~J. Brear, An economic {{MPC}} approach to microgrid
  control, in: 2016 {{Australian Control Conference}} ({{AuCC}}), 2016, pp.
  276--281.
\newblock \href {https://doi.org/10.1109/AUCC.2016.7868202}
  {\path{doi:10.1109/AUCC.2016.7868202}}.

\bibitem{zamani2015integration}
V.~Zamani, A.~Cort{\'e}s, J.~Kleissl, S.~Mart{\'i}nez, Integration of {{PV}}
  generation and storage on power distribution systems using {{MPC}}, in: 2015
  {{IEEE Power}} \& {{Energy Society General Meeting}}, 2015, pp. 1--5.
\newblock \href {https://doi.org/10.1109/PESGM.2015.7286588}
  {\path{doi:10.1109/PESGM.2015.7286588}}.

\bibitem{nair2021analysis}
U.~R. Nair, M.~Sandelic, A.~Sangwongwanich, T.~Dragi{\v c}evi{\'c},
  R.~{Costa-Castell{\'o}}, F.~Blaabjerg, An {{Analysis}} of {{Multi Objective
  Energy Scheduling}} in {{PV-BESS System Under Prediction Uncertainty}}, IEEE
  Transactions on Energy Conversion 36~(3) (2021) 2276--2286.
\newblock \href {https://doi.org/10.1109/TEC.2021.3055453}
  {\path{doi:10.1109/TEC.2021.3055453}}.

\bibitem{pezeshki2014impact}
H.~Pezeshki, P.~J. Wolfs, G.~Ledwich, Impact of {{High PV Penetration}} on
  {{Distribution Transformer Insulation Life}}, IEEE Transactions on Power
  Delivery 29~(3) (2014) 1212--1220.
\newblock \href {https://doi.org/10.1109/TPWRD.2013.2287002}
  {\path{doi:10.1109/TPWRD.2013.2287002}}.

\bibitem{teslapowerwall}
{Tesla}, Tesla {{Powerwall}} 2 {{Datasheet}}, Tech. rep. (Jun. 2019).

\bibitem{TeslasPowerwall2}
Tesla's {{Powerwall}} 2 and {{Solar Roof Tiles}}: {{Our Review}},
  https://unboundsolar.com/solar-information/tesla-powerwall-for-solar.

\bibitem{teslaTESLAPOWERWALLLIMITED2025}
{Tesla},
  \href{https://energylibrary.tesla.com/docs/Public/EnergyStorage/Powerwall/General/Warranty/en-us/Powerwall-Warranty-EN.pdf}{{{TESLA
  POWERWALL LIMITED WARRANTY}} ({{USA}})} (Jan. 2025).
\newline\urlprefix\url{https://energylibrary.tesla.com/docs/Public/EnergyStorage/Powerwall/General/Warranty/en-us/Powerwall-Warranty-EN.pdf}

\bibitem{pgeResidentialRatePlan2024}
{PG\&E}, Electric rates: {{Current}} and historic electric rates,
  https://www.pge.com/tariffs/en/rate-information/electric-rates.html\#accordion-a84c67dc1e-item-e10eec0cc5
  (Apr. 2024).

\bibitem{IEEEGuideLoading2012}
{{IEEE Guide}} for {{Loading Mineral-Oil-Immersed Transformers}} and
  {{Step-Voltage Regulators}}, IEEE Std C57.91-2011 (Revision of IEEE Std
  C57.91-1995) (2012) 1--123\href
  {https://doi.org/10.1109/IEEESTD.2012.6166928}
  {\path{doi:10.1109/IEEESTD.2012.6166928}}.

\bibitem{LarsonElectronics25}
Larson {{Electronics}} - 25 {{KVA Pad Mount Transformer}} - {{12470Y}}/7200
  {{Grounded Wye Primary}}, 240/{{120V Secondary}} - {{Mineral Oil Fluid}} /
  {{ONAN}},
  https://www.larsonelectronics.com/product/292915/25-kva-pad-mount-transformer-12470y-7200-grounded-wye-primary-240-120v-secondary-mineral-oil-fluid-onan.

\bibitem{wilsonEndUseLoadProfiles2022}
E.~Wilson, A.~Parker, A.~Fontanini, E.~Present, J.~Reyna, R.~Adhikari,
  C.~Bianchi, C.~CaraDonna, M.~Dahlhausen, J.~Kim, A.~LeBar, L.~Liu,
  M.~Praprost, L.~Zhang, P.~DeWitt, N.~Merket, A.~Speake, T.~Hong, H.~Li,
  N.~Frick, Z.~Wang, A.~Blair, H.~Horsey, D.~Roberts, K.~Trenbath, O.~Adekanye,
  E.~Bonnema, R.~El~Kontar, J.~Gonzalez, S.~Horowitz, D.~Jones, R.~Muehleisen,
  S.~Platthotam, M.~Reynolds, J.~Robertson, K.~Sayers, Q.~Li, End-{{Use Load
  Profiles}} for the {{U}}.{{S}}. {{Building Stock}}: {{Methodology}} and
  {{Results}} of {{Model Calibration}}, {{Validation}}, and {{Uncertainty
  Quantification}}, Tech. Rep. NREL/TP-5500-80889, 1854582, MainId:78667 (Mar.
  2022).
\newblock \href {https://doi.org/10.2172/1854582} {\path{doi:10.2172/1854582}}.

\bibitem{blonsky2021ochre}
M.~Blonsky, J.~Maguire, K.~McKenna, D.~Cutler, S.~P. Balamurugan, X.~Jin,
  {{OCHRE}}: {{The Object-oriented}}, {{Controllable}}, {{High-resolution
  Residential Energy Model}} for {{Dynamic Integration Studies}}, Applied
  Energy 290 (2021) 116732.
\newblock \href {https://doi.org/10.1016/j.apenergy.2021.116732}
  {\path{doi:10.1016/j.apenergy.2021.116732}}.

\end{thebibliography}

\end{document}